\documentclass[review,hidelinks,onefignum,onetabnum]{siamart251104}
\RequirePackage{packages}

\usepackage{lipsum}
\usepackage{amsfonts}
\usepackage{graphicx}
\usepackage{epstopdf}
\usepackage{algorithmic}
\ifpdf
  \DeclareGraphicsExtensions{.eps,.pdf,.png,.jpg}
\else
  \DeclareGraphicsExtensions{.eps}
\fi

\newsiamremark{remark}{Remark}
\newsiamremark{hypothesis}{Hypothesis}
\crefname{hypothesis}{Hypothesis}{Hypotheses}
\newsiamthm{claim}{Claim}

\headers{Probabilistic Rounding Error Analysis}{S. Bhola, K. Duraisamy}

\title{Bias- and Variance-Aware Probabilistic Rounding Error Analysis for Floating-Point Arithmetic\thanks{Submitted to the editors DATE.
\funding{NSF Grant FMITF-2219997 supported this research}}}

\author{Sahil Bhola\thanks{Department of Aerospace Engineering \& Michigan Institute for Computational Discovery and Engineering,
University of Michigan, Ann Arbor, MI 48109, U.S.A.
  (\email{sbhola@umich.edu}, \email{kdur@umich.edu})}
  \and Karthik Duraisamy\footnotemark[2]
  }

\usepackage{amsopn}

\ifpdf
\hypersetup{
  pdftitle={An Example Article},
  pdfauthor={S. Bhola, K. Duraisamy}
}
\fi

\nolinenumbers
\begin{document}

\maketitle
\begin{abstract}
Probabilistic rounding error analysis can yield much sharper bounds than classical worst-case theory, but existing results typically rely on zero-mean rounding errors and often leave the confidence parameter implicit.
This work revisits probabilistic rounding error analysis in a moment-aware setting.
We first derive a confidence-calibrated reformulation of the Higham and Mary~\cite{higham2019new} bound that makes its confidence parameter explicit.
We then introduce a variance-informed probabilistic backward error bound based on the first two moments of $\log(1+\delta)$, where $\delta$ is the relative rounding error.
This allows the analysis to accommodate biased rounding error models rather than relying on a zero-mean assumption.
To illustrate this framework, we study both a uniform model and a log-space $\operatorname{Beta}$ model for rounding errors, the latter of which provides a simple way to represent bias.
This perspective shows that the growth of probabilistic rounding error bounds is not universal: near-zero-mean regimes recover $\sqrt{n}$-like behavior, while biased models can exhibit faster accumulation.
$\texttt{CUDA}$ experiments in single and half precision on dot products, sparse matrix-vector products, and a stochastic boundary-value problem show that the proposed framework is especially useful in low-precision regimes where deterministic bounds are overly conservative and where bias-aware modeling better matches observed error growth.
\end{abstract}

\begin{keywords}
floating-point arithmetic, probabilistic rounding error analysis, uncertainty quantification
\end{keywords}

\begin{MSCcodes}
65G50, 97N20, 65F99, 65C99
\end{MSCcodes}
\section{Introduction}\label{sec:intro}
Modern computer architectures increasingly support low- or mixed-precision arithmetic to reduce computational complexity, memory access time, and energy footprint.
This has enabled the use of low-precision floating-point formats in a wide-range of applications such as edge computing~\cite{azizi2024low,rachmanto2024characterizing}, deep learning~\cite{gupta2015deep,wang2018training,hubara2018quantized,rokh2023comprehensive}, climate modeling~\cite{hatfield2019accelerating,paxton2022climate,kimpson2023climate}, fluid dynamics~\cite{liu2004real,rinaldi2012lattice,karp2023uncertainty}, and natural sciences~\cite{lienhart2002using}.
Despite the computational advantages, operating in a low-precision format introduces significant rounding errors that can accumulate across successive computations and ultimately degrade accuracy.
This trade-off between efficiency and reliability makes it essential to quantify the effects of rounding errors alongside other numerical errors and statistical uncertainties (e.g., discretization error, parametric uncertainty, and sampling uncertainty).
\emph{Rigorously characterizing rounding error accumulation and deriving reliable error bounds are essential for designing numerical algorithms and statistical models that achieve computational efficiency while retaining provable reliability}.

Traditional rounding error analysis adopts a deterministic worst-case model~\cite{higham1989accurate,higham1990bounding,higham2002accuracy,mori2012backward,kellison2023laproof}, leading to bounds expressed in terms of the operation-count-dependent constant $\gamma_n(\urd) \define n\urd/(1-n\urd)$, where $\urd$ is the unit roundoff and $n$ denotes the number of arithmetic operations with a floating-point number.
Such estimates require $n\urd < 1$ and neglect cancellation effects, making them overly pessimistic when performing a large number of arithmetic operations using low-precision arithmetic~\cite{higham2002accuracy}.
For example, even for simple kernels such as the sequential dot product computed in IEEE half precision, deterministic bounds can overestimate the accumulated rounding errors by several orders of magnitude~\cite{higham2019new}.

To overcome the limitations of deterministic analysis, several works model rounding errors as random variables to capture cancellation effects during successive operations~\cite{von1947numerical,henrici1966test,higham2019new,ipsen2020probabilistic,connolly2023probabilistic}.
In this probabilistic framework, rounding error accumulation is interpreted as uncertainty induced by floating-point arithmetic.
A key heuristic emerging from this viewpoint is that operation-count-dependent constants can often be replaced by their square roots, leading to substantially sharper bounds for algorithms such as dot products, matrix multiplication, and dense linear system solves.
Higham and Mary~\cite{higham2019new} rigorously justified this rule of thumb by modeling rounding errors as independent, zero-mean random variables and applying Hoeffding’s concentration inequality, thereby replacing the classical operation-count-dependent constant $\gamma_n$ with a probabilistic counterpart $\tilde{\gamma}_n$ exhibiting $\sqrt{n}$ growth.
Ipsen and Zhou~\cite{ipsen2020probabilistic} obtained comparable results under the weaker assumption of mean-independence, again assuming zero-mean rounding errors and employing the Azuma--Hoeffding inequality.
Connolly et al.~\cite{connolly2021stochastic} further extended this framework to stochastic rounding.
Higham and Mary~\cite{higham2020sharper} retained the same assumptions on the rounding errors as~\cite{ipsen2020probabilistic}, but additionally assumed independence of the numerical data, enabling sharper bounds when the data distribution has near-zero mean.
Similarly,~\cite{hallman2023precision} analyzed summation over general computational trees and showed that, for tree reductions, sharper bounds can be derived that depend on the tree height instead of the number of inputs for summation.
Collectively, these studies demonstrate that statistically modeling rounding errors yields significantly tighter estimates of the uncertainty due to floating-point arithmetic.
However, existing approaches primarily exploit first-moment information through Hoeffding-type concentration inequalities and typically rely on zero-mean assumptions for rounding errors.
As observed in~\cite{higham2019new}, such assumptions may not hold in practice, limiting the applicability of these probabilistic bounds.

In this work, we develop a general probabilistic rounding error analysis that leverages the \emph{exact first and second moments of the rounding error random variable and applies to arbitrary parameterizations defined in log-space}.
Our work is closely related to the study by El et al.~\cite{el2023stochastic}, which exploits the variance of rounding errors to derive estimates of the uncertainty due to floating-point arithmetic in stochastic rounding.
However, unlike~\cite{el2023stochastic}, our analysis exploits the full characterization of the rounding error random variable rather than relying on bounds on its variance.
Moreover, we do not impose an explicit zero-mean assumption on the rounding errors, and therefore, our framework can accommodate systematic bias via explicit rounding error random variable models.
In summary, the main contributions are as follows:

\begin{enumerate}

    \item \textbf{Variance-informed probabilistic rounding error analysis.}
    We introduce a new operation-count-dependent constant $\hat{\gamma}_n$ that incorporates both the first and second moments of the rounding error random variable, enabling sharper and more flexible quantification of floating-point uncertainty beyond zero-mean assumptions.

    \item \textbf{Explicit and confidence-calibrated probabilistic bounds.}
    We derive a corollary of Theorem 2.4 of Higham and Mary~\cite{higham2019new} that rigorously recovers the $\sqrt{n}$ growth in $\tilde{\gamma}_n$ and provides a closed-form expression for the confidence parameter $\lambda$. Unlike prior formulations where $\lambda$ appears as an arbitrary constant, we express it explicitly in terms of the unit roundoff and required confidence, recovering the scaling $\lambda \propto (1-\urd)^{-1}$ consistent with the empirical findings of Connolly et al.~\cite{connolly2021stochastic}

    \item \textbf{Moment-driven control of accumulation growth.}
    We show that the growth of the operation-count-dependent constant in a probabilistic rounding error analysis stems not merely from stochastic assumptions, but also from how the rounding error distribution is characterized. By modeling bias directly in the log-domain, we demonstrate that the growth of $\hat{\gamma}_n$ can be systematically controlled.

    \item \textbf{GPU-scale numerical validation in low precision.}
    We validate the proposed bounds with $\texttt{CUDA}$ experiments using single precision (\texttt{float}) and half precision (\texttt{\_\_half}) data types on (i) dot products, (ii) sparse matrix-vector multiplication using matrices from the \software{SuiteSparse}~\cite{davis2011university} collection, and (iii) a stochastic ODE where floating-point uncertainty interacts with discretization error and uncertainties from sampling and parameters.
\end{enumerate}

The rest of the paper is organized as follows. In~\S\ref{sec:background}, we review background on floating-point arithmetic, rounding error analysis, and probabilistic bounds. In~\S\ref{sec:methodology}, we present the main theoretical results, including a corollary of Higham and Mary’s analysis that yields explicit confidence-calibrated bounds (\cref{corollary:mprea}) and the variance-informed probabilistic rounding error analysis (\cref{theorem:vprea}). 
In~\S\ref{sec:application}, we apply the framework to quantify uncertainty from floating-point computations in dot products, matrix-vector multiplication, matrix-matrix multiplication, and the solution of a tridiagonal linear system using the Thomas algorithm. Numerical experiments in~\S\ref{sec:experiments} evaluate dot products with random data, matrix-vector multiplication using matrices from the \software{SuiteSparse} collection, and a stochastic boundary value problem, demonstrating the tightness of the proposed bounds. Concluding remarks are presented in~\S\ref{sec:conclusion}.

\noindent {\em Notation:}
Scalars are denoted by lowercase letters (e.g., $\scalar{a}$), vectors by bold lowercase letters (e.g., $\vect{a}$), and matrices by bold uppercase letters (e.g., $\mat{A}$). Matrix and vector elements are denoted by subscripts; for example, $\mat{A}_{i,j}$ denotes the $(i,j)$-th entry.
Inequalities for vectors and matrices hold component-wise; for example, $\abs{\mat{A}} \le \abs{\mat{B}}$ means $\abs{\mat{A}_{i,j}} \le \abs{\mat{B}_{i,j}}$ for all $i,j$. Perturbations are also defined component-wise: $\mat{A}+\Delta\mat{A}$ implies $\mat{A}_{i,j}+\Delta\mat{A}_{i,j}$ for all $i,j$. The set of integers is denoted by $\mathbb{Z}$.

\section{Background}\label{sec:background}
Performing arithmetic operations $op \in \{+,-,\times,/\}$ in floating-point arithmetic introduces rounding errors due to the finite precision of the representation.
For IEEE-754-compliant arithmetic, and in the absence of overflow or underflow, the result of a floating-point arithmetic operation can be modeled as
\begin{align}
\float{x \; op \; y} = (x \; op \; y)(1+\delta)^\rho,
\quad |\delta| \le \urd,
\quad \rho = \pm 1,
\label{eq:ieee_model}
\end{align}
where $\float{\cdot}$ denotes floating-point evaluation and $\urd \define \tfrac{1}{2} r^{1-p}$~\cite{ieee2019ieee,higham2002accuracy}.
Consequently, evaluating a function $\vect{y}=f(\vect{x})$ with $\vect{x}\in\mathbb{F}$ yields a numerical approximation $\hat{\vect{y}} = \hat{f}(\vect{x})$, where $\hat{f}$ denotes the function perturbed by rounding errors introduced at each arithmetic operation.
The accumulation of these errors can be analyzed using either forward or backward error analysis~\cite[Chapter 1]{higham2002accuracy}.
Forward error analysis bounds the deviation between the exact result $\vect{y}$ and the computed result $\hat{\vect{y}}$, whereas backward error analysis seeks the smallest perturbation in the input that explains the computed result.
Specifically, the backward error $\varepsilon_{\text{bwd}}$ is the solution of the minimization problem 
\begin{align}
\varepsilon_{\text{bwd}}
\define
\min \left\{
\varepsilon \ge 0 :
\hat{f}(\vect{x}) = f(\vect{x} + \Delta \vect{x}), \;
\frac{|\Delta \vect{x}|}{|\vect{x}|} \le \varepsilon
\right\}.
\label{eq:backward_error}
\end{align}
In this work, we focus on backward error analysis because it provides a natural framework to compare uncertainty in the input $\vect{x}$ with errors induced by 
floating-point arithmetic and subsequently derive forward error bounds.

In performing backward error analysis using~\cref{eq:ieee_model}, the product $\prod_{i=1}^n(1 + \delta_i)^{\rho_i}$ with $|\delta_i| \leq \urd$ and $\rho_i = \pm 1$ often arises when performing $n$ arithmetic operations with a floating-point number~\cite[Chapter 3]{higham2002accuracy}.
To simplify these product terms and obtain a bound for the backward error, the traditional deterministic rounding error analysis (called \drea, here) assumes the worst-case scenario that only utilizes the bounds for $\delta_i$, as stated in the following lemma~\cite[Lemma 3.1]{higham2002accuracy}.
\begin{lemma}[Deterministic rounding error analysis]
    If $\abs{\delta_i}\le \urd$ and $\rho_i=\pm 1$ for all $i$, and $n\urd<1$, then
\begin{align*}
\prod_{i=1}^n (1+\delta_i)^{\rho_i} = 1 + \theta_n, \qquad \abs{\theta_n} \le \gamma_n(\urd) \define \frac{n\urd}{1-n\urd}.
\end{align*}
\label{lemma:drea}
\end{lemma}
However, such a deterministic approach yields pessimistic estimates of the accumulated error because it cannot account for the cancellation of rounding errors during computation~\cite{higham2019new, croci2022stochastic}.
Further, for $n\urd < 1$, the constant $\gamma_n = \mathcal{O}(n\urd)$ increases linearly with the number of arithmetic operations $n$ and can lead to significant overestimation of the accumulated error when using low-precision arithmetic.

\subsection{Probabilistic Modeling of Rounding Errors}
To account for cancellation of rounding errors during computation, the rounding error can be modeled as a bounded random variable $\delta:\Omega \ra \real{}$, where $\Omega$ denotes the sample space.
\emph{Under this statistical description, quantifying the accumulation of rounding errors can be interpreted as estimating the uncertainty induced by floating-point arithmetic.}
We emphasize that this differs from stochastic rounding, where the rounding direction itself is random~\cite{croci2022stochastic}.
Higham and Mary~\cite[Theorem 2.4]{higham2019new} established the first rigorous result for probabilistic rounding error analysis by modeling rounding errors as independent random variables with zero mean.
They showed that $\gamma_n$ in~\cref{lemma:drea} can be replaced by
\begin{align}
\tilde{\gamma}_n(\urd;\lambda) \define e^{\lambda\sqrt{n}\urd + \frac{n\urd^2}{1-\urd}} -1,
\quad \forall \urd < 1, \ \lambda > 0,
\label{eq:mprea}
\end{align}
such that $\abs{\theta_n} \le \tilde{\gamma}_n(\urd;\lambda)$ holds with probability at least
$\prob{\urd;\lambda} = 1 - 2 e^{-\frac{\lambda^2(1-\urd)^2}{2}}$.
Here $\lambda$ is a hyperparameter that controls the confidence level of the bound.
We refer to this result as \emph{Mean-informed Probabilistic Rounding Error Analysis} (\mprea).

Unlike deterministic analysis, the probabilistic formulation remains valid for any number of operations $n$.
However, it introduces additional modeling assumptions.
In particular, Higham and Mary introduce the \(\sqrt{n}\) scaling in~\cref{eq:mprea} to make the associated probability measure independent of \(n\), an assumption that is not intrinsic to floating-point arithmetic itself but imposed for analytical tractability.
Moreover, the bound depends only on the first moment of the rounding error distribution due to the use of Hoeffding's concentration inequality~\cite{boucheron2003concentration}.
Finally, the analysis assumes that rounding errors have zero mean, an assumption that may not hold in practice~\cite{higham2019new}.

\begin{lemma} [Hoeffding's Inequality]
    Let $\{Z_i\}_{i=1}^n$ be $n$ independent random variables with $\abs{Z_i} \leq c_i$ for all $i$. Then, the random variable $S_n \define \sum_{i=1}^n Z_i$ satisfies
\begin{equation*}
    \prob{\abs{S_n - \expect{S_n}} \geq t} \leq 2 e^{\frac{-t^2}{2\sum_{i=1}^n c_i^2}} ,
\end{equation*}
for all positive $t$.
\label{lemma:hoeffding_inequality}
\end{lemma}

\section{Methodology}\label{sec:methodology}
In this section, we first present a corollary of the probabilistic rounding error analysis of Higham and Mary~\cite{higham2019new} that (a) rigorously establishes the $\sqrt{n}$ dependence in $\tilde{\gamma}_n$ and (b) improves the interpretability of $\mprea$ by deriving a closed-form expression for $\lambda$ in~\cref{eq:mprea} in terms of the desired confidence level.
We then introduce the central contribution of this work, \emph{Variance-informed Probabilistic Rounding Error Analysis} ($\vprea$), which leverages both the first and second moments of the rounding error random variable to quantify uncertainty induced by floating-point arithmetic.
We show that \emph{$\vprea$ applies to any independent and identically distributed rounding error random variable, provided closed-form expressions for the first two moments of $\log(1+\delta)$ are available.}

\subsection{An Interpretable Mean-informed Probabilistic Rounding Error Analysis}
We now present a corollary of Theorem $2.4$ in~\cite{higham2019new} that introduces an exact functional form of $\lambda$ and the rounding error bounds in terms of the required confidence.
We show that $\lambda$ depends on the unit roundoff and derive the $\lambda \propto ( 1 - \urd)^{-1}$, as empirically found by Connolly et al.~\cite{connolly2021stochastic}.
The goal is to improve the interpretability of probabilistic rounding error bounds by eliminating unnecessary modeling assumptions.

\begin{corollary}[Mean-informed probabilistic rounding error analysis]\label{corollary:mprea}
    Let $\{\delta_i\}_{i=1}^n$ be $n$ independent random variables with zero mean with $\abs{\delta_i} \leq \urd < 1$ for all $i$, where $\urd$ is the unit roundoff.
    Then, for $\rho_i = \pm 1$, the relation
    \begin{align*}
        \prod_{i=1}^n (1+\delta_i)^{\rho_i} = 1 + \theta_n, \qquad \abs{\theta_n} \le \tilde{\gamma}_n(\urd;\zeta) \define e^{t + \frac{n\urd^2}{1 - \urd}} - 1,
    \end{align*}
    holds with confidence $\zeta\in[0, 1)$, where $t \define \frac{\urd}{1 - \urd} \sqrt{ - 2n \log\Big(\frac{1-\zeta}{2}\Big)}$.
    \begin{proof}
        Consider the random variable $\log\phi\define \sum_{i=1}^n \rho_i\log(1 + \delta_i)$ that is the sum of $n$ independent random variables $\rho_i\log(1 + \delta_i)$.
        Then, for $\abs{\delta_i}\leq \urd < 1$ for all $i$, we can use the Taylor series expansion $\log(1 + \delta_i) = \sum_{k=1}^\infty \frac{(-1)^{k+1}\delta_i^k}{k}$ to obtain the bounds
        \begin{align*}
        \delta_i - \frac{\delta_i^2}{1 - \abs{\delta_i}} \leq \log(1 + \delta_i) \leq \delta_i + \frac{\delta_i^2}{1 - \abs{\delta_i}}.
        \end{align*}
        Therefore, $\abs{\rho_i \log(1 + \delta_i)} = \abs{\log(1 + \delta_i)} \leq \urd + \frac{\urd^2}{1 - \urd} = \frac{\urd}{1 - \urd}.$
        Since $\expect{\delta_i} = 0$, we have $\abs{\expect{\log(1 + \delta_i)}} \leq \expect{\abs{\log(1 + \delta_i)}} \leq \expect{\abs{\delta_i + \frac{\delta_i^2}{1 - \abs{\delta_i}}}} \leq \frac{\urd^2}{1 - \urd}$.
        Now, we can use~\cref{lemma:hoeffding_inequality} with $Z_i = \rho_i \log(1 + \delta_i)$ and $c_i = \frac{\urd}{1 - \urd}$ to obtain
        \begin{align*}
            \prob{\abs{\log\phi - \expect{\log\phi}} \geq t} \leq 2 e^{\frac{-t^2}{2nc^2}},\\
            \prob{\abs{\log\phi - \expect{\log\phi}} \leq t} \geq 1 - 2 e^{ \frac{-t^2}{2nc^2}},
        \end{align*}
        where $c \define \frac{\urd}{1 - \urd}$.
        Let $\abs{\log\phi - \expect{\log\phi}} \leq t$ hold with confidence $\zeta\in[0, 1)$.
        For a confidence $\zeta$, we can now obtain the value of the distance parameter $t$ by solving the quadratic
        \begin{align*}
            t^2 + 2nc^2 \log\Big(\frac{1 - \zeta}{2}\Big) = 0.
        \end{align*}
        The roots of this quadratic are given as
        \begin{align*}
            t_{-} \define - c \sqrt{ 2n \log\Big(\frac{2}{1-\zeta}\Big)}, \qquad t_{+} \define c \sqrt{ 2n \log\Big(\frac{2}{1 - \zeta}\Big)}.
        \end{align*}
        Since $t_- < 0$ for all $n > 0$, $\urd < 1$, and $\zeta \in (0, 1]$, the inequality $\abs{\log\phi - \expect{\log\phi}} \leq t_-$ will never hold true.
        Similarly, $t_+ > 0$ for all $n > 0$, $\urd < 1$, and $\zeta \in (0, 1]$, such that $\abs{\log\phi - \expect{\log\phi}} \leq t_+$ holds with confidence at least $\zeta$.

        Now, we can invoke the inequality $\abs{\log\phi - \expect{\log\phi}} \geq \abs{\log\phi} - \abs{\expect{\log\phi}} \geq \abs{\log\phi} - \frac{n\urd^2}{1 - \urd}$ to obtain
        \begin{align*}
            \prob{\abs{\log\phi} - \frac{n\urd^2}{1 - \urd} \leq t_+} \geq \prob{\abs{\log\phi - \expect{\log\phi}} \leq t_+} \geq \zeta,
        \end{align*}
        such that $\prob{\abs{\log\phi} \leq t_+ + \frac{n\urd^2}{1 - \urd}} \geq \zeta$.
        Thus, $\log\phi \in [-(t_+ + \frac{n\urd^2}{1 - \urd}), (t_+ + \frac{n\urd^2}{1 - \urd})]$ with confidence at least $\zeta$, which gives
        \begin{align*}
            \phi - 1 \in [e^{-(t_+ + \frac{n\urd^2}{1 - \urd})} - 1, e^{(t_+ + \frac{n\urd^2}{1 - \urd})} - 1].
        \end{align*}
        Therefore, we can obtain $\abs{\phi - 1} \leq \max\{\abs{e^{-(t_+ + \frac{n\urd^2}{1 - \urd})} - 1}, \abs{e^{(t_+ + \frac{n\urd^2}{1 - \urd})} - 1}\} = e^{(t_+ + \frac{n\urd^2}{1 - \urd})} - 1 = \tilde{\gamma}_n(\urd;\zeta)$,
        that holds with confidence at least $\zeta$.
    \end{proof}
\end{corollary}
    Note that the probabilistic analysis of Higham and Mary assumes the specific functional form \(t = \lambda \sqrt{n}\,\urd\) in order to eliminate the explicit dependence of the probability measure on \(n\).
    More generally, for any scalar-valued function \(g(\urd;\lambda)\), setting \(t = \sqrt{n}\, g(\urd;\lambda)\) would preserve this independence and leave the analysis formally unchanged.
    Consequently, the particular choice of functional form is not uniquely determined by the theory and is instead a modeling assumption.
    On the contrary, we show that the functional form of $t$ can be obtained rigorously by defining a confidence parameter $\zeta\in[0, 1)$ with which the rounding error bounds are satisfied.
    Comparing the expression of $t$ in~\cref{corollary:mprea} with $\lambda\sqrt{n}\urd$, we can identify an exact functional form $\lambda(\urd;\zeta) = \frac{1}{1 - \urd}\sqrt{2\log\Big(\frac{2}{1-\zeta}\Big)}$,
    thus rigorously obtaining the $\lambda \propto (1 - \urd)^{-1}$ dependence as empirically found by Connolly et al.~\cite{connolly2021stochastic}.

\subsection{Variance-informed Probabilistic Rounding Error Analysis}
We now introduce a new approach to probabilistic rounding error analysis that uses the first and second moments of the rounding error random variable.
To do so, we leverage Bernstein's concentration inequality~\cite{boucheron2003concentration} as stated in the following lemma.
\begin{lemma} [Bernstein's Inequality]
    Let $\{Z_i\}_{i=1}^n$ be $n$ independent random variables such that $\abs{Z_i}\leq c $ for all $i$.
Then, the random variable $S_n \define \sum_{i=1}^n Z_i$ satisfies
\begin{align*}
    \prob{\abs{S_n - \expect{S_n}} \geq t} \leq 2 \exp \Big ( \frac{-t^2}{2(\sigma^2 + \frac{ct}{3})}\Big),
\end{align*}
for all positive $t$, where $\sigma^2 \define \sum_{i=1}^n \var{Z_i}$.
\label{lemma:bernstein_inequality}
\end{lemma}

\begin{theorem}[Variance-informed probabilistic rounding error analysis]\label{theorem:vprea}
    Let $\{\delta_i\}_{i=1}^n$ be $n$ independent and identically distributed random variables with $\abs{\delta_i} \leq \urd < 1$ for all $i$, where $\urd$ is the unit roundoff.
    Then, for $\rho_i=\pm 1$, the relation
    \begin{align*}
        \prod_{i=1}^{n} (1 + \delta_i)^{\rho_{i}} = 1 + \theta_n, \qquad \abs{\theta_n} \leq \hat{\gamma}_n(\urd;\zeta) \define e^{t + n\abs{\hat{\mu}}} - 1,
    \end{align*}
    holds with confidence $\zeta\in[0, 1)$, where
    \begin{align*}
        t \define \frac{1}{3} \left(c \log\left(\frac{2}{1-\zeta}\right) + \sqrt{\left(\frac{\urd}{1-\urd} \log\left(\frac{1-\zeta}{2}\right)\right)^2 - 18n \log\left(\frac{1-\zeta}{2}\right) \hat{\sigma}^2}\right),
    \end{align*}
    with $\hat{\mu}\define \expect{\log(1 + \delta_i)}$, $\hat{\sigma}^2 \define \var{\log(1 + \delta_i)}$, and $\abs{\log(1 + \delta_i)} \leq c$ for all $i$.
    \begin{proof}
        Consider the random variable $\log\phi\define \sum_{i=1}^n \rho_i \log(1 + \delta_i)$ that is the sum of $n$ independent and identically distributed random variables $\rho_i \log(1 + \delta_i)$.
        We can then apply~\cref{lemma:bernstein_inequality} with $Z_i = \rho_i \log(1 + \delta_i)$ to obtain
        \begin{align*}
            \prob{\abs{\log(\phi) - \expect{\log(\phi)}} \geq t} \leq 2 \exp \Big ( \frac{-t^2}{2(n\hat{\sigma}^2 + \frac{ct}{3})}\Big),\\
            \prob{\abs{\log(\phi) - \expect{\log(\phi)}} \leq t} \geq 1 - 2 \exp \Big ( \frac{-t^2}{2(n\hat{\sigma}^2 + \frac{ct}{3})}\Big),
        \end{align*}
        where $\sigma^2 = n \hat{\sigma}^2 = \var{\log(1 + \delta_i)}$ for independent and identically distributed random variables, and $c$ denotes the bound on $\abs{\log(1 + \delta_i)}$ for all $i$.
        Let $\abs{\log(\phi) - \expect{\log(\phi)}} \leq t$ hold with confidence $\zeta\in[0, 1)$.
        For such confidence $\zeta$, we can now obtain the value of the distance parameter $t$ by solving the quadratic
        \begin{align*}
            t^2 + \frac{2ct}{3} \log\Big(\frac{1 - \zeta}{2}\Big) + 2n\hat{\sigma}^2 \log\Big(\frac{1 - \zeta}{2}\Big) = 0.
        \end{align*}
        The roots of this quadratic are
        \begin{align*}
         t_{\pm} = \frac{1}{3}\left( - c \log\!\left(\frac{1-\zeta}{2}\right) \pm \sqrt{ \left(c \log\!\left(\frac{1-\zeta}{2}\right)\right)^2 - 18 n \log\!\left(\frac{1-\zeta}{2}\right)\hat{\sigma}^2 } \right),
        \end{align*}
        where $t_+$ and $t_-$ correspond to the positive and negative branches, respectively.
        For positive $\hat{\sigma}, n, $ and $\urd$, both $t_-$ and $t_+$ are real-valued and distinct for all $\zeta \in [0, 1)$.
        Under such a condition for $\hat{\sigma}, n ,$ and $\urd$, we will always have a negative root $t_-$ and a positive root $t_+$ for all $\zeta \in [0, 1)$.
        Thus, the inequality $\abs{\log(\phi) - \expect{\log(\phi)}} \leq t_-$ will never hold true.
        Therefore, using the positive root $t_+$, the inequality $\abs{\log(\phi) - \expect{\log(\phi)}} \leq t_+$ holds with confidence at least $\zeta$.

        Now, we can invoke the inequality $\abs{\log(\phi) - \expect{\log\phi}} \geq \abs{\log\phi} - \abs{\expect{\log\phi}} \geq \abs{\log\phi} - n \abs{\hat{\mu}}$, where $\hat{\mu}\define \expect{\log(1 + \delta_i)}$, to obtain
        \begin{align*}
            \prob{\abs{\log\phi} - n \abs{\hat{\mu}} \leq t_+} \geq \prob{\abs{\log\phi - \expect{\log\phi}} \leq t_+} \geq \zeta,
        \end{align*}
        such that $\prob{\abs{\log\phi} \leq t_+ + n \abs{\hat{\mu}}} \geq \zeta$.
        Thus, $\log\phi \in [-(t_+ + n \abs{\hat{\mu}}), (t_+ + n \abs{\hat{\mu}})]$ with confidence at least $\zeta$, which gives
        \begin{align*}
            \phi - 1 \in [e^{-(t_+ + n \abs{\hat{\mu}})} - 1, e^{(t_+ + n \abs{\hat{\mu}})} - 1].
        \end{align*}
        Therefore, we can obtain $\abs{\phi - 1} \leq \max\{\abs{e^{-(t_+ + n \abs{\hat{\mu}})} - 1}, \abs{e^{(t_+ + n \abs{\hat{\mu}})} - 1}\} = e^{(t_+ + n \abs{\hat{\mu}})} - 1 = \hat{\gamma}_n(\urd;\zeta)$, that holds with confidence at least $\zeta$.
    \end{proof}
\end{theorem}

Note that $\hat{\gamma}_{n+1} > \hat{\gamma}_n$ for all positive $n$, that is, the rounding error bounds monotonically increase with the number of arithmetic operations.
\Cref{theorem:vprea} presents a general framework for quantifying uncertainty due to floating-point arithmetic when the statistics of the rounding error random variables are known and finite.
It generalizes~\cref{corollary:mprea} by utilizing both first and second moments of the rounding error random variable, with the additional assumption that the rounding errors are identically distributed.
Furthermore, the proposed approach does not require an explicit assumption about the moments of the rounding error random variable, unlike~\cref{corollary:mprea}, which requires $\expect{\delta}=0$.

\subsubsection{Modeling rounding error random variable}
To quantify the uncertainty due to floating-point arithmetic using~\cref{theorem:vprea}, we introduce two models for the rounding error random variable $\delta$.
The motivation of these models is (a) to remain agnostic to the underlying floating-point operation, (b) to obtain closed-form moments required in~\cref{theorem:vprea}, and (c) to account for potential bias in rounding errors.

\begin{figure}[h!]
    \centering
    \includegraphics[width=0.72\textwidth]{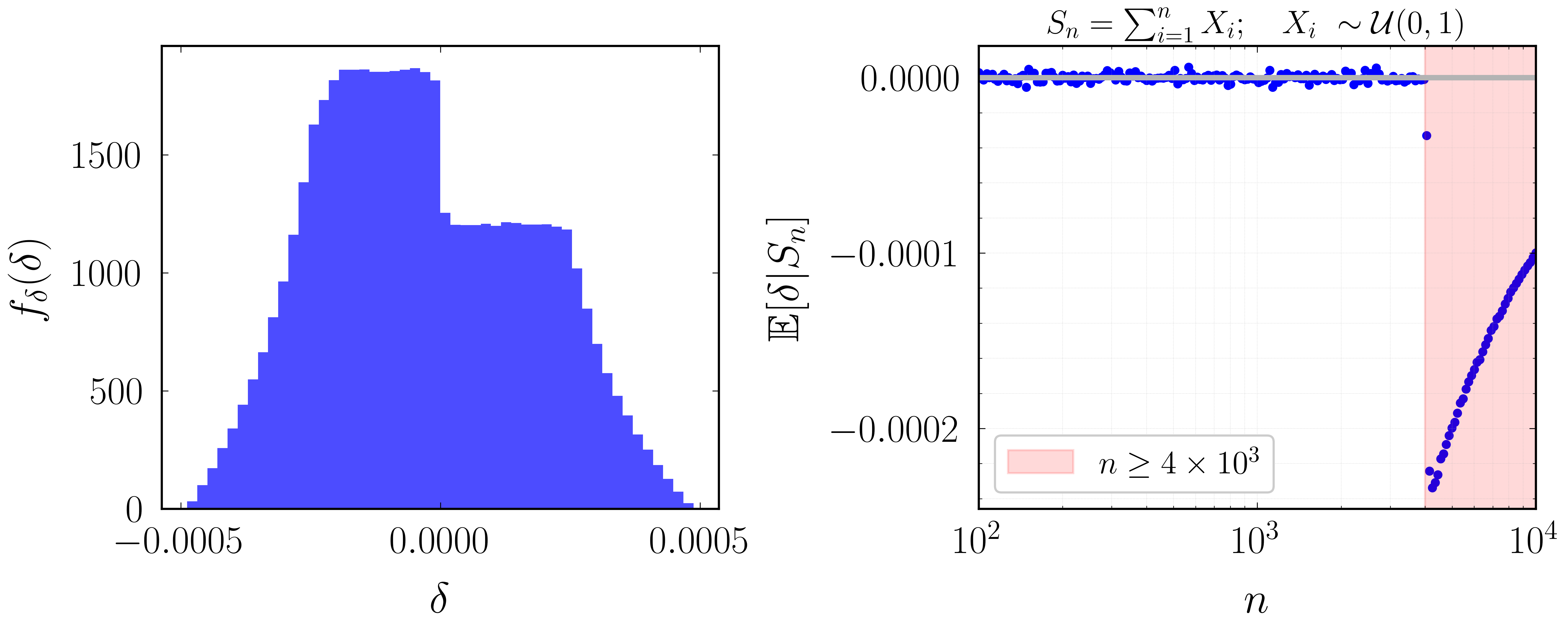}
    \caption{Empirical distribution (left) and conditional expectation (right) of rounding error random variable $\delta = \frac{\float{ S_{n} + X_{n+1} } - (S_n+X_{n+1})}{(S_n+X_{n+1})}$, where $S_n = \sum_{i=1}^n X_i$ and $X_i~\sim\mathcal{U}(0, 1)$ for all $i$.
    Here, computations are performed using half-precision floating-point arithmetic.
    To obtain the statistics, $10^4$ independent experiments were conducted for each $n$.
    }
    \label{fig:rounding_error_distribution_small_increments}
\end{figure}

First, we consider an uninformative uniform distribution as outlined in~\cref{def:uniform_rounding_error_model}.
\begin{definition}[$\mathcal{U}$-model]
    The rounding errors due to finite-precision floating point arithmetic is modeled as an independent and identically distributed random variable with an uninformative uniform distribution $\delta\sim\mathcal{U}(-\urd, \urd)$.
    \label{def:uniform_rounding_error_model}
\end{definition}
Under~\cref{def:uniform_rounding_error_model}, we can obtain
\begin{subequations}\label{eq:uniform_moments}
\begin{align}
    \hat{\mu} &= \frac{-2u + (-1 + u) \log(1 - u) + (1 + u) \log(1 + u)}{2u},\\
    \hat{\sigma}^2 &= \frac{4u^2 + \kappa [ \log^2(1 - u) -2 \log(1-u)\log(1+u) + \log^2(1 + u)]}{4u^2},\\
    c  &= \log(1 + \urd),
\end{align}
\end{subequations}
as defined in~\cref{theorem:vprea}, where $\kappa \define (-1 + \urd^2)$.
An obvious limitation of such a model is that it may not accurately quantify uncertainty due to floating-point arithmetic when rounding errors are biased, that is, $\expect{\delta}\neq 0$.
This arises because~\cref{def:uniform_rounding_error_model} implicitly enforces $\expect{\delta} = 0$, thereby embedding a zero-mean assumption on the rounding error, similar to~\cref{corollary:mprea}, where this requirement is imposed explicitly.
However, such an assumption, implicit or explicit, may not always hold true in practical computations.
For example, when adding a very large positive number with a very small positive number, the rounding errors tend to be negatively biased, as shown in~\cref{fig:rounding_error_distribution_small_increments}.

To account for the potential bias in rounding errors distribution, we also consider a $\operatorname{Beta}$ distribution to model $\log(1+\delta)$ random variable, as outlined in~\cref{def:beta_rounding_error_model}.

\begin{definition}[$\beta$-model]
Let $\delta$ denote the rounding error arising from finite-precision floating-point arithmetic, and define the random variable $Y = \log(1+\delta)$.
We model $Y$ as an independent and identically distributed random variable of the form
\begin{align*}
Y = \log(1-\mathrm{u}) + \log\!\left(\frac{1+\mathrm{u}}{1-\mathrm{u}}\right)\, Z, \qquad Z \sim \operatorname{Beta}(\alpha,\beta),
\end{align*}
    where $\mathrm{u}$ denotes the unit roundoff, and $\alpha,\beta\ireal{}_{>0}$ are the shape parameters of the $\operatorname{Beta}$ distribution.
    \label{def:beta_rounding_error_model}
\end{definition}
Under~\cref{def:beta_rounding_error_model}, we can obtain
\begin{subequations}\label{eq:beta_moments}
\begin{align}
    \hat{\mu} &= \log(1 - \urd) + \log\Big(\frac{1 + \urd}{1 - \urd}\Big) \frac{\alpha}{\alpha + \beta},\\
    \hat{\sigma}^2 &= \Big(\log\Big(\frac{1 + \urd}{1 - \urd}\Big)\Big)^2 \frac{\alpha\beta}{(\alpha + \beta)^2 (\alpha + \beta + 1)},\\
    c &= \log(1 + \urd),
\end{align}
\end{subequations}
as defined in~\cref{theorem:vprea}.
Using~\cref{def:beta_rounding_error_model}, we can bias rounding errors by appropriately selecting the shape parameters $\alpha$ and $\beta$, as formalized in the following proposition.
\begin{proposition}\label{prop:beta_positive_mean}
    Let $Z \sim \operatorname{Beta}(\alpha,\beta)$ and define
    $Y = \log(1-\urd) + \log(\ell) Z$, where
    $\ell \define \frac{1+\urd}{1-\urd} > 1$, and
    $h(\urd) \define -\log(1-\urd)/\log(\ell)$.
    Then the expectation of the rounding error random variable $\delta$ is strictly positive for all $\alpha > \frac{h(\urd)}{1-h(\urd)}\,\beta$ and is strictly negative for all $\alpha < \frac{h(\urd)}{1 - h(\urd)} \,\beta$.
\begin{proof}
    Using~\cref{def:beta_rounding_error_model}, we have $\delta = e^Y - 1$ that is convex function.
    Thus, using Jensen's inequality, we have $\expect{\delta} \geq e^{\expect{Y}} - 1$.
    We can now obtain the condition for $\expect{\delta} > 0$ by requiring $\expect{Y} > 0$.
    Using~\cref{eq:beta_moments}, we have $\expect{\delta} > 0$ when $\frac{\alpha}{\alpha + \beta} > h(\urd)$, which gives the condition $\alpha > \frac{h(\urd)}{1-h(\urd)}\,\beta$.
    Now, we can obtain the condition for making $\expect{\delta} < 0$.
    Using the definition of $Y$, we can obtain $\delta = (1-\urd) \ell^Z - 1$.
    To obtain a strictly negative expectation, we require $(1-\urd) \expect{\ell^Z}  < 1$.
    Note, for $\ell > 1$ and $Z \sim \operatorname{Beta}(\alpha,\beta)$, $\ell^Z$ is a convex function with $\expect{\ell^Z} \geq \ell^{\expect{Z}}$ via Jenson's inequality.
    Thus, we can obtain the condition of making $\expect{\delta} < 0$ as
    \begin{align*}
        \expect{Z} \log(\ell) \leq \log(\expect{\ell^Z}) < -\log(1-\urd),
    \end{align*}
    which gives the condition $\frac{\alpha}{\alpha + \beta} < h(\urd)$, or equivalently, $\alpha < \frac{h(\urd)}{1-h(\urd)}\,\beta$.
\end{proof}
\end{proposition}
\begin{figure}
    \centering
    \includegraphics[width=0.4\textwidth]{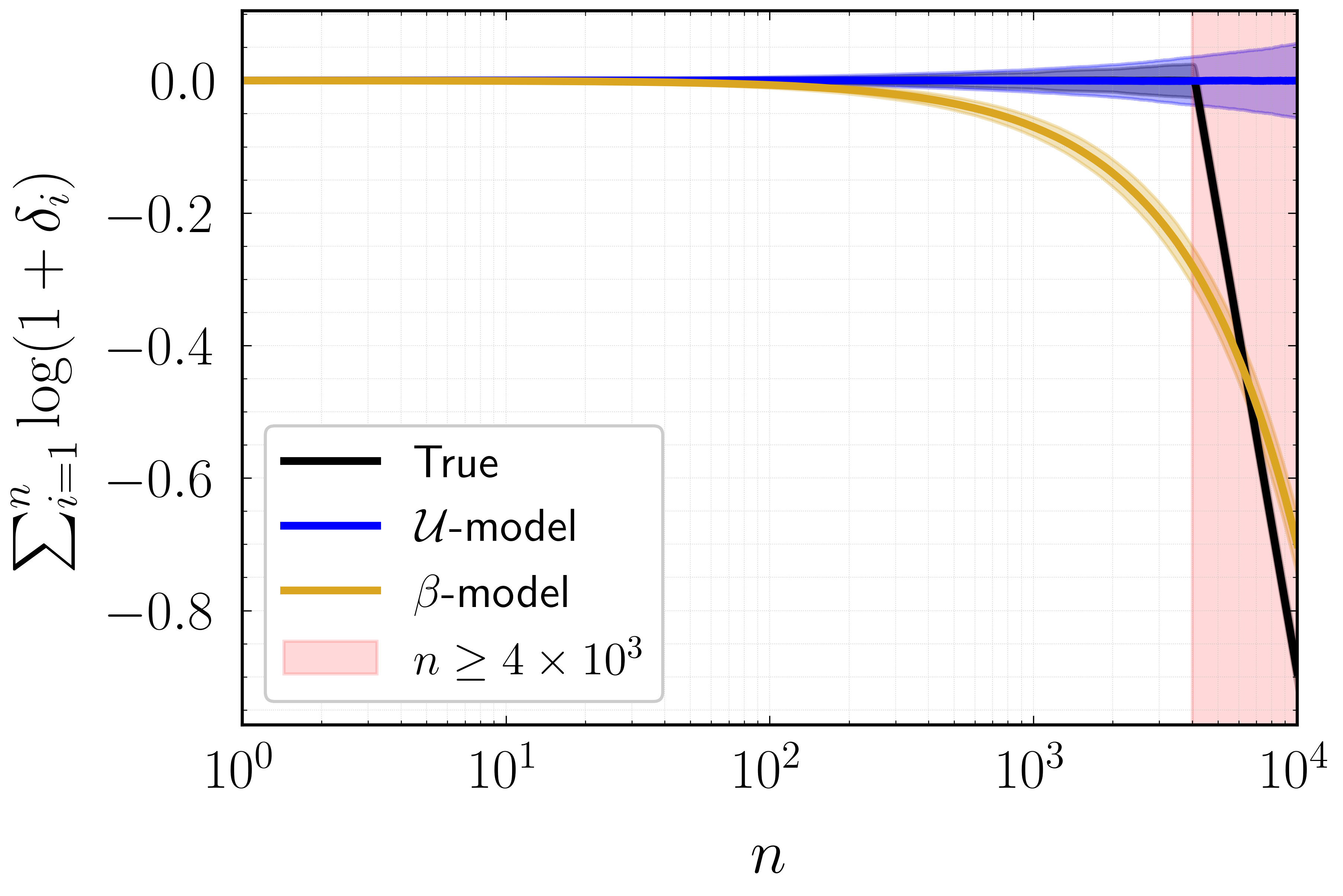}
    \caption{An illustration of the random walk $\sum_{i=1}^{n}\log(1+\delta_i)$, where $\delta = \frac{\float{ S_{n} + X_{n+1} } - (S_n+X_{n+1})}{(S_n+X_{n+1})}$ with $S_n = \sum_{i=1}^n X_i$ and $X_i~\sim\mathcal{U}(0, 1)$ for all $i$.
    Here, computations are performed using half-precision floating-point arithmetic.
    To obtain the statistics, $10^4$ independent trajectories were computed, with solid lines denoting the sample mean and shaded regions denoting two standard deviations about the mean.
    For the $\beta$-model (\cref{def:beta_rounding_error_model}), we used shape parameters $\alpha = 1.5$ and $\beta = 2.0$. }
    \label{fig:random_walk_of_product}
\end{figure}

\begin{figure}[h!]
    \centering
    \includegraphics[width=0.82\textwidth]{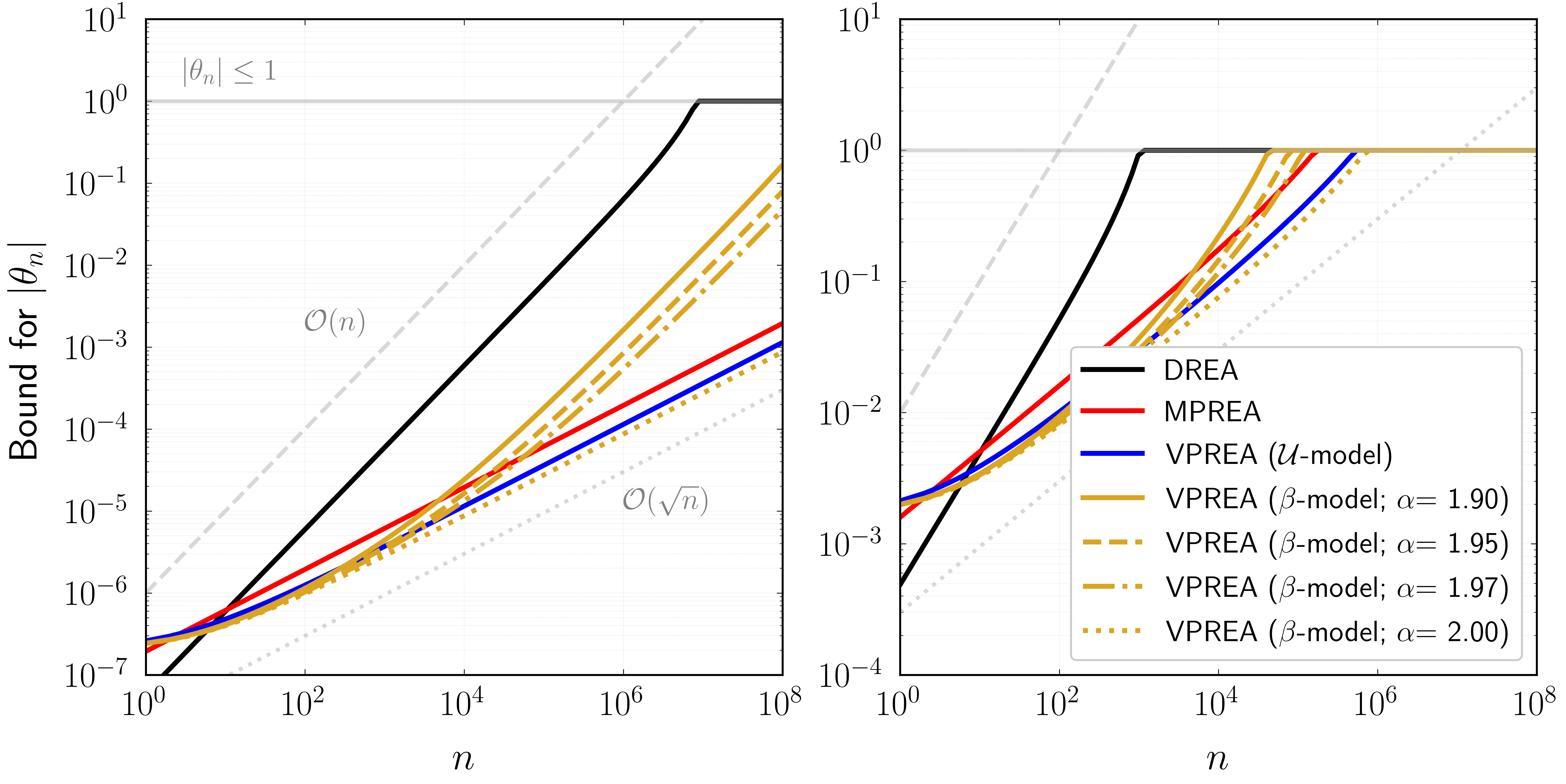}
    \caption{ Comparison of rounding error bounds obtained using \drea (\textcolor{black}{\fline}), \mprea (\textcolor{red}{\fline}), and \vprea under the $\mathcal{U}$-model (\textcolor{blue}{\fline}) and the $\beta$-model (\textcolor{goldenrod}{\fline,\dashed,\dotted}).
    Results are shown for single-precision (left) and half-precision (right) floating-point arithmetic.
    All probabilistic bounds are evaluated using a confidence level $\zeta = 0.99$, and the $\beta$-model uses the shape parameter $\beta = 2.0$.
    We choose the shape parameter $\alpha$ such that $\expect{\delta}$ is strictly negative, thus accounting for the negative bias in rounding errors observed when adding small increments to a large sum, as shown in~\cref{fig:rounding_error_distribution_small_increments}.
    Here, we plot $\gamma_n$, $\tilde{\gamma}_n$, and $\hat{\gamma}_n$ until they exceed one (\textcolor{refgrey}{\fline}), beyond which they are not meaningful for backward error analysis.
    }
    \label{fig:gamma_vs_n}
\end{figure}

Thus, by appropriately selecting the shape parameters $\alpha$ and $\beta$, we can model rounding error random variables with positive or negative bias using~\cref{def:beta_rounding_error_model}.
As illustrated in~\cref{fig:random_walk_of_product}, the $\beta$-model can capture the negative bias in the rounding error random variable when adding small increments to a large sum, as observed in~\cref{fig:rounding_error_distribution_small_increments}.

\Cref{fig:gamma_vs_n} illustrates the bounds for $\abs{\theta_n}$ obtained using \drea, \mprea, and \vprea using the $\mathcal{U}$-model and the $\beta$-model for IEEE single and half precision floating-point arithmetic.
Compared to \drea that scales as $\mathcal{O}(n)$, both \mprea and \vprea scale slowly with the number of arithmetic operations and produce smaller bounds for $n\gtrsim10$.
Furthermore, probabilistic approaches can produce meaningful bounds (less than one for an informative backward error analysis) for a significantly larger number of arithmetic operations than \drea.
As observed, \vprea ($\mathcal{U}$-model) grows similar to \mprea as $\mathcal{O}(\sqrt{n})$ since both models assume rounding error random variables have zero mean, either in an implicit or explicit manner.
However, by incorporating the second central moment of the rounding error random variable, \vprea\ ($\mathcal{U}$-model) attains the same scaling as \mprea, with a smaller multiplicative constant.
Consequently, the resulting bounds are smaller for nearly all $n$, except for a very small regime $n \lesssim 5$, where the bounds are marginally larger.
Using the $\beta$-model for \vprea, we can introduce a negative bias in the rounding error random variable (using~\cref{prop:beta_positive_mean}), and therefore can parameterize the growth of the bounds via the shape parameters of the $\operatorfont{Beta}$ distribution.
As shown, for the shape parameters $\beta = 2.0$ and $\alpha = 2.00$ (for which $\expect{\delta}$ is marginally negative), the bound exhibits $\mathcal{O}(\sqrt{n})$ growth.
Decreasing the value of the shape parameter $\alpha$ (thereby increasing the negative bias in the rounding error random variable) results in a faster growth of the bounds with $n$, transitioning between $\mathcal{O}{(\sqrt{n})}$ growth and $\mathcal{O}{(n)}$ growth rate.

\section{Uncertainty due to floating-point arithmetic in numerical linear algebra kernels}\label{sec:application}
Using~\cref{theorem:vprea}, we now derive statistical bounds to quantify uncertainty arising from floating-point arithmetic in three fundamental kernels: (a) dot products, (b) matrix-vector products, and (c) the Thomas algorithm for solving tridiagonal linear systems.
These operations form the computational backbone of many scientific computing tasks, including matrix factorizations and linear systems arising from numerical discretizations.
In particular, the Thomas algorithm is widely used for banded systems encountered in fluid dynamics, structural dynamics, and control applications.
Reliable estimates of the uncertainty due to floating-point arithmetic enable the safe use of low-precision arithmetic in large-scale simulations and support the development of low-computational-fidelity solvers within multi-fidelity statistical frameworks.
Since practical interest lies in the accumulated rounding error incurred by representing the statistical model in floating-point arithmetic, we assume identical precision for inputs and computations.
For probabilistic rounding error analysis, we define
\begin{align*}
\mathcal{Q}(n;\zeta) \define 1 - n(1 - \zeta),
\end{align*}
which follows from the inclusion-exclusion principle and yields a lower bound on the probability that all $n$ bounds hold simultaneously, provided each bound holds with probability at least $\zeta \in [0,1)$.

\subsection{Dot product}\label{sec:la_dp}
In the following, we present the statistical bounds for the uncertainty due to floating-point arithmetic when performing a dot product of two vectors using~\cref{theorem:vprea}.
\begin{theorem}[Dot product]\label{theorem:vprea_dot_product}
    Let $y = \vect{a}^T\vect{b}$, where $\vect{a}, \vect{b} \ireal{n}$.
    Then, the computed dot product $\hat{y}$ satisfies
    \begin{align*}
        \hat{y} = (\mathbf{a} + \Delta \mathbf{a})^T \mathbf{b} = \mathbf{a}^T (\mathbf{b} + \Delta \mathbf{b}),
    \end{align*}
    where $\abs{\Delta \mathbf{a}} \leq \hat{\gamma}_{n}\abs{\vect{a}}$ and $\abs{\Delta \mathbf{b}} \leq \hat{\gamma}_{n}\abs{\vect{b}}$ hold true with a probability of at least $\mathcal{Q}(n;\zeta)$.
    \begin{proof}
        Let us assume that $y = s_n$, where $s_i = s_{i-1} + \vect{a}_i \vect{b}_i$ is a recursive relation with $s_0 = 0$.
        Here $\vect{a}_i$ and $\vect{b}_i$ are the $i$-th elements of $\vect{a}$ and $\vect{b}$, respectively.
        Then, the computed dot product $\hat{y}$ can be represented as
        \begin{align*}
            \hat{y} &= \sum_{i=1}^{n} \vect{a}_i\vect{b}_i(1+\eta_i) \prod_{j = \max\{2, i\}}^n (1+\xi_j),\\
            &= \sum_{i=1}^n \vect{a}_i\vect{b}_i (1 + \theta_{n-\max\{2, i\}+2}),
        \end{align*}
    where $\eta_i$ and $\xi_i$ are the rounding errors due to multiplication and addition, respectively, such that $\abs{\eta_i}, \abs{\xi_i} \leq \urd$.
        We can invoke~\cref{theorem:vprea} to write the computed dot product as using perturbed vectors as
    \begin{align*}
        \hat{y} &= \sum_{i=1}^n (\vect{a}_i + \Delta \vect{a}_i) \vect{b}_i = \sum_{i=1}^n \vect{a}_i (\vect{b}_i + \Delta \vect{b}_i),
    \end{align*}
    where $\Delta \vect{a}_i \define \vect{a}_i \theta_{n - \max\{2, i\} + 2}$ and $\Delta \vect{b}_i \define \vect{b}_i \theta_{n - \max\{2, i\} + 2}$, such that,
    \begin{align*}
    \abs{\Delta \vect{a}_i} \leq \hat{\gamma}_{n - \max\{2, i\} + 2} \abs{\vect{a}_i} \leq \hat{\gamma}_n \abs{\vect{a}_i},\\
    \abs{\Delta \vect{b}_i} \leq \hat{\gamma}_{n - \max\{2, i\} + 2} \abs{\vect{b}_i} \leq \hat{\gamma}_n \abs{\vect{b}_i},\\
    \end{align*}
    is satisfied with a probability of at least $\zeta$ and will fail with at most $1 - \zeta$.
    Thus, for all $i$, the bounds will hold with a probability of at least $\mathcal{Q}(n;\zeta)$ using the principles of inclusion and exclusion of probabilities.
    \end{proof}
\end{theorem}

Using~\cref{theorem:vprea_dot_product}, we can obtain the estimate for uncertainty due to floating-point arithmetic in computing the dot product as
\begin{align}
    \frac{\abs{\hat{y} - y}}{\abs{y}} \leq  \hat{\gamma}_n \frac{\abs{\vect{a}^T}\abs{\vect{b}}}{\abs{\vect{a}^T\vect{b}}},
\end{align}
that holds true with a probability of at least $\mathcal{Q}(n;\zeta)$.

\subsection{Matrix-vector product}\label{sec:matrix_multiply}
Using~\cref{theorem:vprea_dot_product}, we can now derive the statistical bounds for the uncertainty due to floating-point arithmetic in computing matrix-vector products.

\begin{theorem}[Matrix-vector product]\label{theorem:vprea_matrix_vector}
    Let $\vect{y} = \mat{a}\vect{x}$, where $\mat{a}\ireal{m\times n}, \vect{x}\ireal{n}$ and $\vect{y}\ireal{m}$.
    Then, the computed matrix-vector product $\hat{\vect{y}}$ satisfies
    \begin{align*}
        \hat{\vect{y}} = (\mat{a} + \Delta \mat{a}) \vect{x},
    \end{align*}
    where $\abs{\Delta\mat{a}} \leq \hat{\gamma}_n\abs{\mat{A}}$ holds true with a probability of at least $\mathcal{Q}(mn;\zeta)$.
    \begin{proof}
        Let $\vect{y}_i$ and $\vect{a}_i$ denote the $i$-th element of $\vect{y}$ and the $i$-th row of $\mat{a}$, respectively.
        Then, we can write $\vect{y}_i = \vect{a}_i^T \vect{x}$ for all $i$.
        Using~\cref{theorem:vprea_dot_product}, we can now write the $i$-th element of the computed matrix-vector product $\hat{\vect{y}}$ as
        \begin{align*}
            \hat{\vect{y}}_i &= (\vect{a}_i + \Delta \vect{a}_i)^T \vect{x},
        \end{align*}
        where $\abs{\Delta \vect{a}_i} \leq \hat{\gamma}_n \abs{\vect{a}_i}$ holds true with a probability of at least $\mathcal{Q}(n;\zeta)$.
        Note that, to compute all the elements of $\hat{\vect{y}}$, we need to perform $m$ dot products, each of size $n$.
        Thus, for all $i$, the bounds will hold true with a probability of at least $\mathcal{Q}(mn;\zeta)$ using the principles of inclusion and exclusion of probabilities.
    \end{proof}
\end{theorem}

Using~\cref{theorem:vprea_matrix_vector}, we can obtain the estimate for uncertainty due to floating-point arithmetic in computing matrix-vector products as
\begin{align}
    \frac{\abs{\hat{\vect{y}} - \vect{y}}}{\abs{\vect{y}}} \leq  \hat{\gamma}_n \frac{\abs{\mat{a}}\abs{\vect{x}}}{\abs{\mat{a}\vect{x}}},
    \label{eq:matrix_vector_product_forward_error_bound}
\end{align}
that holds true with a probability of at least $\mathcal{Q}(mn;\zeta)$.
We can similarly derive the statistical bounds for matrix-matrix products using~\cref{theorem:vprea_matrix_vector} by considering each column of the resulting matrix as a matrix-vector product.

\begin{theorem}\label{theorem:vprea_matrix_matrix}
    Let $\mat{C} = \mat{A}\mat{B}$, where $\mat{A}\ireal{m\times t}, \mat{B}\ireal{t\times n}$ and $\mat{C}\ireal{m\times n}$.
    Then, the computed matrix-matrix product $\hat{\mat{C}}$ satisfies
    \begin{align*}
        \frac{\abs{\hat{\mat{C}} - \mat{C}}}{\abs{\mat{C}}} \leq \hat{\gamma}_t \frac{\abs{\mat{A}}\abs{\mat{B}}}{\abs{\mat{A}\mat{B}}},
    \end{align*}
    that holds true with a probability of at least $\mathcal{Q}(mnt;\zeta)$.
    \begin{proof}
        Let $\vect{c}_i$ and $\vect{b}_i$ denote the $i$-th column of $\mat{C}$ and $\mat{B}$, respectively.
        Then, we can write $\vect{c}_i = \mat{A}\vect{b}_i$ for all $i$.
        Using~\cref{theorem:vprea_matrix_vector}, we can now write the $i$-th column of the computed matrix-matrix product $\hat{\mat{C}}$ as
        \begin{align*}
            \hat{\vect{c}}_i &= (\mat{A} + \Delta \mat{A}) \vect{b}_i,
        \end{align*}
        where $\abs{\Delta \mat{A}} \leq \hat{\gamma}_t \abs{\mat{A}}$ holds true with a probability of at least $\mathcal{Q}(mt;\zeta)$.
        Note that, to compute all the columns of $\hat{\mat{C}}$ we need to perform $n$ matrix-vector products.
        Thus, for all $i$, the bound $\abs{\hat{\vect{c}}_i - \vect{c}_i} \leq \hat{\gamma}_t \abs{\mat{a}}\abs{\vect{b}_i}$ holds true with a probability of at least $\mathcal{Q}(mnt;\zeta)$ using the principles of inclusion and exclusion of probabilities.
    \end{proof}
\end{theorem}

\subsection{Thomas algorithm for solving a tri-diagonal linear system}\label{sec:thomas}
Thomas' algorithm is a specialized method for solving a linear system $\mat{a}\vect{x}=\vect{b}$ via LU-factorization when the coefficient matrix $\mat{a}\ireal{n\times n}$ is tridiagonal~\cite{trefethen2022numerical}.
Broadly, it consists of three sequential steps, namely (a) LU-factorization, (b) forward substitution, and (c) backward substitution.
As a result, to quantify the uncertainty due to floating-point arithmetic in the Thomas algorithm, we must first individually quantify the uncertainty accumulated at each of the outlined steps.

\paragraph{LU-factorization}
Consider the tridiagonal linear system $\mat{A}\vect{x}=\mat{b}$, where $\mat{A}\in\mathbb{R}^{n\times n}$ has subdiagonal, diagonal, and superdiagonal entries denoted by $\alpha_i$, $\beta_i$, and $\nu_i$ at the $i$-th row, respectively.
Applying Doolittle’s method~\cite{higham2002accuracy}, the matrix $\mat{A}$ admits an LU-actorization $\mat{A}=\mat{L}\mat{U}$, where $\mat{L}\in\mathbb{R}^{n\times n}$ is unit lower triangular and $\mat{U}\in\mathbb{R}^{n\times n}$ is upper triangular.
The nonzero entries of $\mat{L}$ and $\mat{U}$ are given by
\begin{subequations}\label{eq:lu_tridiagonal}
\begin{align}
    \mat{L}_{i,i-1} &\define l_i = \frac{\alpha_i}{\mat{U}_{i-1, i-1}},
    \quad && i = 2,\dots,n,\label{eq:lu_tridiagonal_lower}\\
    \mat{U}_{i,i}   &\define u_i = \beta_i - \mat{L}_{i, i-1} \nu_{i-1},
\quad && i = 1,\dots,n, \label{eq:lu_tridiagonal_upper}
\end{align}
\end{subequations}
with $\mat{L}_{i,i}=1$ for all $i$, $\mat{U}_{i,i+1}=\nu_i$ for $i=1,\dots,n-1$, and $\mat{u}_{1, 1}=\beta_1$.
In the following lemma, we present the statistical bounds for the uncertainty due to floating-point arithmetic in the LU-factorization of a tridiagonal matrix via Doolittle's method.
\begin{lemma}[LU-factorization of a tri-diagonal system]\label{lemma:lu}
    Let $\mat{a}=\mat{lu}$ be the $LU$-factorization of a tri-diagonal matrix $\mat{A}\ireal{n\times n}$ that has $\alpha_i$, $\beta_i$, and $\nu_i$ as its subdiagonal, diagonal, and superdiagonal entries at the $i$-th row, respectively.
    Then, the computed factorization $\hat{\mat{l}}\ireal{n\times n}$ and $\hat{\mat{u}}\ireal{n\times n}$ via Doolittle's method satisfies
    \begin{align*}
        \hat{\mat{L}}\hat{\mat{U}} = \mat{A} + \Delta \mat{A},
    \end{align*}
    where $\abs{\Delta\mathbf{A}} \leq \hat{\gamma}_1\hat{\abs{\mathbf{L}}}\hat{\abs{\mathbf{U}}}$ holds true with a probability of at least $\mathcal{Q}(3(n-1);\zeta)$.
    \begin{proof}
        Since the arithmetic operations involved in computing the LU-factorization of a tri-diagonal matrix are given by~\cref{eq:lu_tridiagonal}, the computed nonzero entries $\hat{l}_i$ and $\hat{u}_i$ are given as
        \begin{align*}
            \hat{l}_i &= \frac{\alpha_i}{\hat{u}_{i-1}(1+\eta_i)}, \quad && i=2,\dots,n,\\
            \hat{u}_i &= \frac{\left(\beta_i - \hat{l}_i \nu_{i-1}(1+\psi_i)\right)}{(1+\xi_i)}, \quad && i=2,\dots,n,
        \end{align*}
        where $\eta_i$ and $\psi_i$ denote the rounding errors arising from multiplications, and $\xi_i$ denotes the rounding error arising from additions.
        Here, $\abs{\eta_i}, \abs{\psi_i}, \abs{\xi_i} \leq \urd$.
        We can now invoke~\cref{theorem:vprea} to express the computed entries $\hat{l}_i$ and $\hat{u}_i$ as
        \begin{align*}
            \hat{l}_i\hat{u}_{i-1} + \theta_1 \hat{l}_i u_{i-1} &= \alpha_i, \quad && i=2,\dots,n,\\
            \hat{u}_i + \hat{l}_i \nu_{i-1} + \theta_1^\prime\hat{u}_i + \theta_1^{\prime\prime}\hat{l}_i \nu_{i-1} &= \beta_i, \quad && i=2,\dots,n,
        \end{align*}
        where $\abs{\theta}_1, \abs{\theta_1^\prime}, \abs{\theta_1^{\prime\prime}} \leq \hat{\gamma}_1$ hold true with a probability of at least $\zeta$.
        Thus, for the $i$-th row of the computed LU-factorization, all the bounds are satisfied with a probability of at least $\mathcal{Q}(3;\zeta)$ using the principles of inclusion and exclusion of probabilities.
        Since we have $n-1$ such computations, all the bounds will hold true with a probability of at least $\mathcal{Q}(3(n-1);\zeta)$.
    \end{proof}
\end{lemma}

\paragraph{Forward substitution}
Once the LU-factorization of the tridiagonal matrix $\mat{A}$ is computed, we can perform forward substitution to solve the system $\mat{L}\vect{y}=\vect{b}$.
In the following lemma, we present statistical bounds for the uncertainty due to floating-point arithmetic during forward substitution in the Thomas algorithm.
\begin{lemma}[Forward substitution in Thomas algorithm]\label{lemma:forward_subs}
    Let $\mat{L}\vect{y} = \vect{b}$ be a tri-diagonal system where $\mat{L}\ireal{n\times n}$ is the lower-triangular matrix obtained from the LU-factorization of a tri-diagonal matrix $\mat{A}\ireal{n\times n}$.
    Then, the computed solution $\hat{\vect{y}}$ via forward substitution satisfies
    \begin{align*}
        (\mat{L} + \Delta \mat{L}) \hat{\vect{y}} = \vect{b},
    \end{align*}
    where $\abs{\Delta \mat{L}}\leq \hat{\gamma}_1\abs{\mat{L}}$ holds true with a probability of at least $\mathcal{Q}(2(n-1);\zeta)$.
    \begin{proof}
        Consider the lower-triangular system $\mat{L}\vect{y}=\vect{b}$, where the nonzero entries of $\mat{L}$ are given by~\cref{eq:lu_tridiagonal_lower}.
        Thus, the solution $\vect{y}$ using forward substitution is given via the recursive relation
        \begin{align*}
            \vect{y}_i = \vect{b}_i - l_i \vect{y}_{i-1}, \quad && i=2,\dots,n,
        \end{align*}
        where $\vect{y}_1 = \vect{b}_1$ and $l_i$ is defined in~\cref{eq:lu_tridiagonal_lower}.
        Thus, we can obtain the computed solution $\hat{\vect{y}}$ as
        \begin{align*}
            \hat{\vect{y}}_i(1+\xi_i) = \vect{b}_i - \hat{l}_i \hat{\vect{y}}_{i-1}(1+\eta_i), \quad && i=2,\dots,n,
        \end{align*}
        where $\eta_i$ and $\xi_i$ are the rounding errors due to multiplication and addition, respectively, such that $\abs{\eta_i}, \abs{\xi_i} \leq \urd$.
        We can now invoke~\cref{theorem:vprea} to express the computed solution as
        \begin{align*}
            \hat{\vect{y}}_i + l_i \hat{\vect{y}}_{i-1} + \theta_1 \hat{\vect{y}}_i + \theta_1^\prime l_i \hat{\vect{y}}_{i-1} = \vect{b}_i, \quad && i=2,\dots,n,
        \end{align*}
        where $\abs{\theta_1}, \abs{\theta_1^\prime} \leq \hat{\gamma}_1$ hold true with a probability of at least $\zeta$.
        Thus, for the $i$-th row of the forward substitution, all the bounds are satisfied with a probability of at least $\mathcal{Q}(2;\zeta)$ using the principles of inclusion and exclusion of probabilities.
        Since we have $n-1$ such computations, all the bounds will hold true with a probability of at least $\mathcal{Q}(2(n-1);\zeta)$.
    \end{proof}
\end{lemma}

\paragraph{Backward substitution}
Once the forward substitution is performed, we can perform backward substitution to solve the system $\mat{U}\vect{x}=\vect{y}$ to obtain the solution of the tri-diagonal system $\mat{a}\vect{x}=\vect{b}$.
In the following lemma, we present statistical bounds on the uncertainty arising from floating-point arithmetic during backward substitution in the Thomas algorithm.
\begin{lemma}[Backward substitution in Thomas algorithm]\label{lemma:backward_subs}
    Let $\mat{U}\vect{x} = \vect{y}$ be a tri-diagonal system where $\mat{U}\ireal{n\times n}$ is the upper-triangular matrix obtained from the LU-factorization of a tri-diagonal matrix $\mat{A}\ireal{n\times n}$.
    Then, the computed solution $\hat{\vect{x}}$ via backward substitution satisfies
    \begin{align*}
        (\mat{U} + \Delta \mat{U}) \hat{\vect{x}} = \vect{y},
    \end{align*}
    where $\abs{\Delta \mat{U}}\leq \hat{\gamma}_2\abs{\mat{U}}$ holds true with a probability of at least $\mathcal{Q}(2n-1;\zeta)$.
    \begin{proof}
        Consider the upper-triangular system $\mat{U}\vect{x}=\vect{y}$, where the nonzero entries of $\mat{U}$ are given by~\cref{eq:lu_tridiagonal_upper}.
        Thus, the solution $\vect{x}$ using backward substitution is given via the recursive relation
        \begin{align*}
            \vect{x}_i = \frac{\vect{y}_i - \nu_i \vect{x}_{i+1}}{u_i}, \quad && i=n-1,\dots,1,
        \end{align*}
        where $\vect{x}_n = \frac{y_n}{u_n}$ and $u_i$ is defined in~\cref{eq:lu_tridiagonal_upper}.
        Thus, we can obtain the computed solution $\hat{\vect{x}}$ as
        \begin{align*}
        \hat{\vect{x}}_n
        &= \frac{y_n}{u_n (1+\eta_n)}, \\
        \hat{\vect{x}}_i
        &= \frac{\vect{y}_i - \nu_i\hat{\vect{x}}_{i+1}(1+\psi_i)} {u_i (1+\xi_i)(1+\chi_i)}, \qquad i = n-1, \ldots, 1 .
        \end{align*}
    where $\eta_i$, $\psi_i$, and $\xi_i$ denote the rounding errors arising from multiplications, and $\chi_i$ denotes the rounding error arising from additions.
    Here, $\abs{\eta_i}, \abs{\psi_i}, \abs{\xi_i}, \abs{\chi_i} \leq \urd$.
    We can now invoke~\cref{theorem:vprea} to express the computed solution $\hat{\vect{x}}$ as
    \begin{align*}
        u_n \hat{\vect{x}}_n + \theta_1 u_n \hat{\vect{x}}_n = \vect{y}_n, \\
        u_i \hat{\vect{x}}_i + \nu_i \hat{\vect{x}}_{i+1} + \theta_2 u_i \hat{\vect{x}}_i + \theta_1^\prime \nu_i \hat{\vect{x}}_{i+1} = \vect{y}_i, \quad && i=n-1,\dots,1,
    \end{align*}
    where $\abs{\theta_1}, \abs{\theta_1^\prime}, \abs{\theta_2} \leq \hat{\gamma}_2$ hold true with a probability of at least $\zeta$.
    Thus, for the $n$-th row of the backward substitution, all the bounds are satisfied with a probability of at least $\mathcal{Q}(1;\zeta)$ using the principles of inclusion and exclusion of probabilities.
    Similarly, for any $i < n$, the bounds are satisfied with a probability of at least $\mathcal{Q}(2;\zeta)$.
    Therefore, all the bounds will hold true with a probability of at least $\mathcal{Q}(2n-1;\zeta)$.
    \end{proof}
\end{lemma}

\paragraph{Thomas algorithm}
Using~\cref{lemma:lu,lemma:forward_subs,lemma:backward_subs}, we can now obtain the statistical bounds for the uncertainty due to floating-point arithmetic in solving a tri-diagonal system of equations via the Thomas algorithm.
We present this in the following theorem.
\begin{theorem}[Thomas algorithm]\label{theorem:vprea_thomas}
    Let $\mat{A}\vect{x}=\vect{b}$ be a tri-diagonal system with $\mat{A}\ireal{n\times n}, \vect{x}\ireal{n}$ and $\vect{b}\ireal{n}$.
    Assume that $\hat{\mat{L}}$ and $\hat{\mat{U}}$ are the computed LU-factorization of the matrix $\mat{A}$ via Doolittle's method.
    Then, the computed solution $\hat{\vect{x}}$ satisfies
    \begin{align*}
        (\mathbf{A} + \Delta \mathbf{A}) \hat{\mathbf{x}} = \mathbf{b},
    \end{align*}
    where $\abs{\Delta \mathbf{A}} \leq (2\hat{\gamma}_1 + \hat{\gamma}_2 + \hat{\gamma}_1\hat{\gamma}_2) \abs{\hat{\mathbf{L}}}\abs{\hat{\mathbf{U}}}$ holds true with a probability of at least $\mathcal{Q}(7n - 6;\zeta)$.
    \begin{proof}
        Consider the computed $LU$-factorization \((\hat{\mat{L}},\hat{\mat{U}})\) of the tridiagonal matrix \(\mat{A}\) satisfying \(\hat{\mat{L}}\hat{\mat{U}}=\mat{A}+\Delta\mat{A}_1\), where \(\abs{\Delta\mat{A}_1}\le \hat{\gamma}_1\abs{\hat{\mat{L}}}\abs{\hat{\mat{U}}}\) holds with probability at least $\mathcal{Q}(3(n-1);\zeta)$, as established in~\cref{lemma:lu}.
        Now, using~\cref{lemma:forward_subs}, we have the computed solution $\hat{\vect{y}}$ of the system $\hat{\mat{L}}\vect{y}=\vect{b}$ satisfying $(\hat{\mat{L}}+\Delta\hat{\mat{L}})\hat{\vect{y}}=\vect{b}$, where $\abs{\Delta\hat{\mat{L}}}\le \hat{\gamma}_1\abs{\hat{\mat{L}}}$ holds with probability at least $\mathcal{Q}(2(n-1);\zeta)$.
        Finally, using~\cref{lemma:backward_subs}, we have the computed solution $\hat{\vect{x}}$ of the system $\hat{\mat{U}}\vect{x}=\hat{\vect{y}}$ satisfying $(\hat{\mat{U}}+\Delta\hat{\mat{U}})\hat{\vect{x}}=\hat{\vect{y}}$, where $\abs{\Delta\hat{\mat{U}}}\le \hat{\gamma}_2\abs{\hat{\mat{U}}}$ holds with probability at least $\mathcal{Q}(2n-1;\zeta)$.
        Combining these results, we obtain
        \begin{align*}
            (\hat{\mat{L}} + \Delta\hat{\mat{L}})(\hat{\mat{U}} + \Delta\hat{\mat{U}})\hat{\vect{x}} = \vect{b}, \\
            (\hat{\mat{L}}\hat{\mat{U}} + \hat{\mat{L}}\Delta\hat{\mat{U}} + \Delta\hat{\mat{L}}\hat{\mat{U}} + \Delta\hat{\mat{L}}\Delta\hat{\mat{U}})\hat{\vect{x}} = \vect{b}, \\
            (\mat{A} + \Delta\mat{A}_1 + \hat{\mat{L}}\Delta\hat{\mat{U}} + \Delta\hat{\mat{L}}\hat{\mat{U}} + \Delta\hat{\mat{L}}\Delta\hat{\mat{U}})\hat{\vect{x}} = \vect{b},\\
            (\mat{A} + \Delta\mat{A})\hat{\vect{x}} = \vect{b},
        \end{align*}
        where $\Delta\mat{A} \define \Delta\mat{A}_1 + \hat{\mat{L}}\Delta\hat{\mat{U}} + \Delta\hat{\mat{L}}\hat{\mat{U}} + \Delta\hat{\mat{L}}\Delta\hat{\mat{U}}$, such that $\abs{\mat{a}} \leq (2\hat{\gamma}_1 + \hat{\gamma}_2 + \hat{\gamma}_1\hat{\gamma}_2) \abs{\hat{\mat{l}}}\abs{\hat{\mat{u}}}$ holds true with a probability of at least $\mathcal{Q}(7n - 6;\zeta)$
        via the principles of inclusion and exclusion of probabilities.
    \end{proof}
\end{theorem}

Using~\cref{theorem:vprea_thomas}, we can now obtain the estimate for uncertainty due to floating-point arithmetic in solving a tri-diagonal system of equations as
\begin{align}
    \frac{\abs{\hat{\vect{x}} - \vect{x}}}{\abs{\vect{x}}} \leq \frac{(2\hat{\gamma}_1 + \hat{\gamma}_2 + \hat{\gamma}_1\hat{\gamma}_2) \abs{\mat{a}^{-1}} \abs{\mat{a}} \abs{\hat{\vect{x}}}}{\abs{\vect{x}}},
\end{align}
that holds true with a probability of at least $\mathcal{Q}(7n - 6;\zeta)$.

\section{Numerical experiments}\label{sec:experiments}
In this section, we present numerical experiments to evaluate the proposed probabilistic rounding error analysis framework.
We compare our approach with the probabilistic method of Higham and Mary~\cite{higham2019new} (see~\cref{corollary:mprea}) and the worst-case deterministic analysis in~\cref{lemma:drea}.
We first examine statistical bounds for rounding error uncertainty in dot products and matrix-vector products with random data.
To reflect practical settings in which multiple error sources coexist, we also study a stochastic boundary-value problem that incorporates parameter uncertainty, sampling variability, floating-point arithmetic, and discretization error.
All experiments are implemented in $\texttt{C++}$ with $\texttt{CUDA~12.5}$ on an $\text{A}100$ GPU.
Reference solutions are computed using IEEE double-precision arithmetic using the standard $\texttt{double}$ type.
Half precision uses the $\texttt{CUDA}$ $\texttt{\_\_half}$ type, while single precision uses the standard $\texttt{float}$ type.


\subsection{Dot product}\label{sec:experiment_dp}
\begin{figure}
    \centering
    \includegraphics[width=0.9\textwidth]{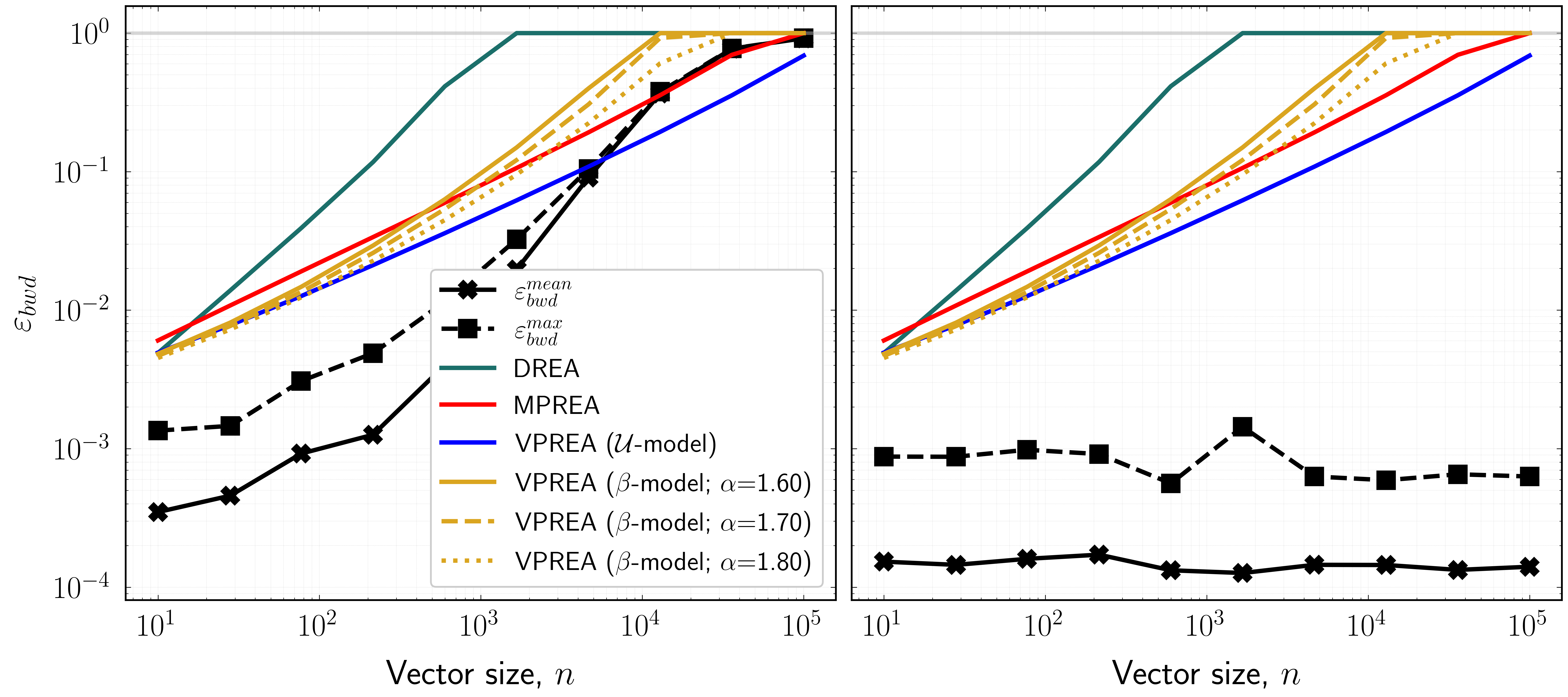}
    \caption{Backward error and its bounds for the dot product of random vectors of size $n$ distributed as $\mathcal{U}(0,1)$ (left) and $\mathcal{U}(-1,1)$ (right), computed using half-precision arithmetic.
    All probabilistic bounds are evaluated using a confidence level $\mathcal{Q}(n;\zeta) = 0.99$, and the $\beta$-model uses the shape parameter $\beta = 2.0$.
    To obtain the statistics, $10^2$ independent experiments were conducted for each $n$.
    All bounds are plotted until they exceed one (\textcolor{refgrey}{\fline}), beyond which they are not meaningful for backward error analysis.
    }
    \label{fig:dot_product_backward_error_half}
\end{figure}

Consider the dot product $y = \vect{a}^T \vect{b}$ of two vectors $\vect{a},\vect{b}\ireal{n}$, for which the backward error using~\cref{eq:backward_error} is given as
\begin{align}
\varepsilon_{bwd}  = \min \left\{ \varepsilon \geq 0 : \hat{y} = (\vect{a}+\Delta \vect{a})^T\vect{b}, \; \dfrac{\abs{\Delta \vect{a}}}{\abs{\vect{a}}} \leq \varepsilon \right\} = \frac{\abs{\hat{y} - y}}{\abs{\vect{a}}^T\abs{\vect{b}}},
\end{align}
where $\hat{y}$ is the computed dot product using floating-point arithmetic.
Similar to~\cite{higham2019new}, we consider the data distributions $\mathcal{U}(0, 1)$ and $\mathcal{U}(-1, 1)$ for the random vectors.
This choice is motivated to study the sharpness of the bounds in two complementary regimes: (a) when adding small positive numbers to an already large partial sum, which is known to produce negative mean rounding error, and (b) when the data distribution admits both positive and negative values, allowing for cancellation effects in the dot product computation.
\begin{figure}
    \centering
    \includegraphics[width=0.9\textwidth]{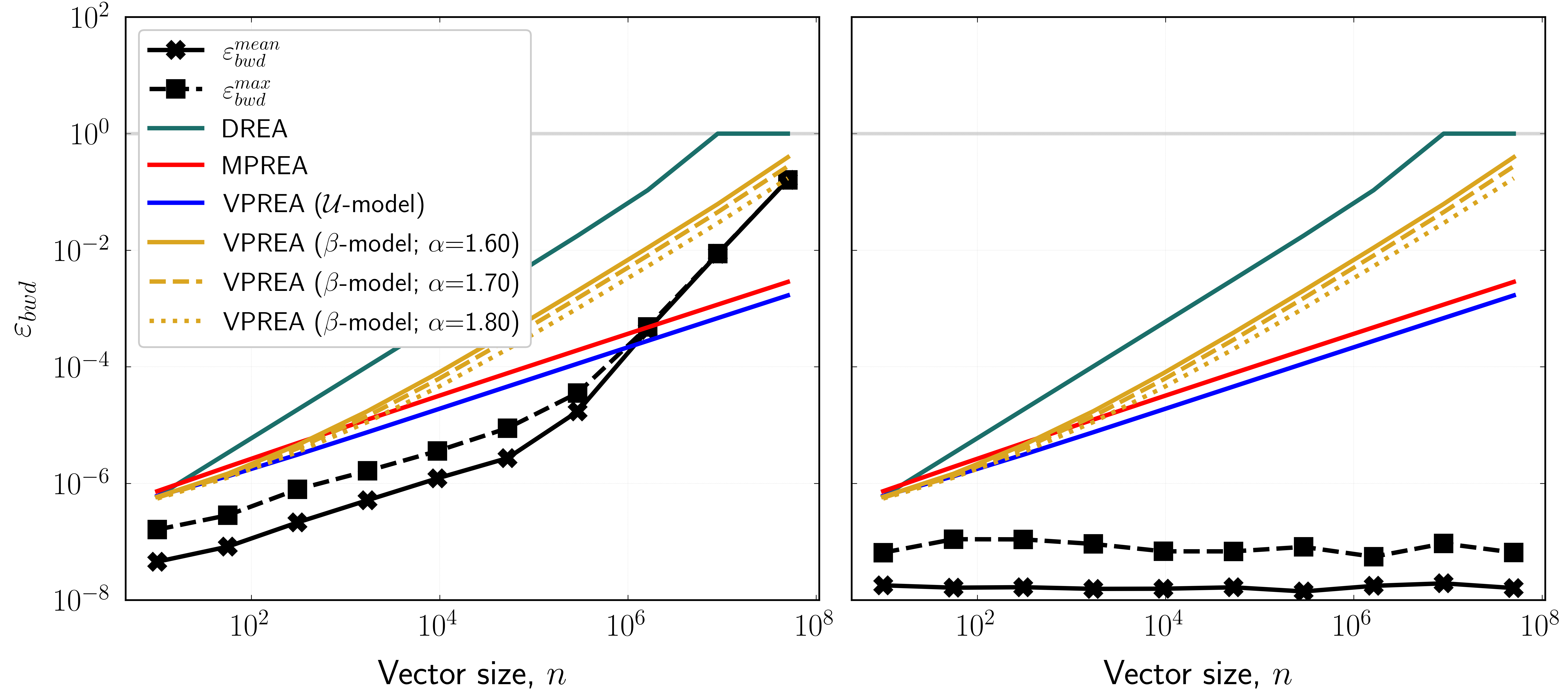}
    \caption{Backward error and its bounds for the dot product of random vectors of size $n$ distributed as $\mathcal{U}(0,1)$ (left) and $\mathcal{U}(-1,1)$ (right), computed using single-precision arithmetic.
    All probabilistic bounds are evaluated using a confidence level $\mathcal{Q}(n;\zeta) = 0.99$, and the $\beta$-model uses the shape parameter $\beta = 2.0$.
    To obtain the statistics, $10^2$ independent experiments were conducted for each $n$.
    All bounds are plotted until they exceed one (\textcolor{refgrey}{\fline}), beyond which they are not meaningful for backward error analysis.
    }
    \label{fig:dot_product_backward_error_single}
\end{figure}
\Cref{fig:dot_product_backward_error_single} illustrates the backward error and its bounds (using~\cref{theorem:vprea_dot_product}) for the dot product of random vectors of size $n$ computed using single-precision arithmetic.
As shown, \vprea provides a more accurate estimate of the backward error than \mprea and \drea for a wide range of vector sizes, especially in the case of $\mathcal{U}(0,1)$ data distribution where rounding errors can potentially have negative expectations (as discussed in~\cref{sec:methodology}).
When the data is distributed as $\mathcal{U}(-1,1)$, all bounds are extremely pessimistic as they do not assume any specific prior information about the data distribution and thereby are unable to account for cancellation in dot product computation.
This, however, motivates normalizing the data to have a zero mean to improve computational accuracy in floating-point arithmetic.
As expected, \vprea ($\mathcal{U}$-model) scales similar to \mprea, however, both bounds begin to fail for sufficiently large $n$ as they cannot admit a negative expectation of the rounding error random variable.
In this regime, \vprea ($\beta$-model) is able to provide a better estimate of backward error growth by accounting for the negative expectation of rounding errors via~\cref{prop:beta_positive_mean}.
Consequently, it provides improvements over \drea for a wide range of $n$, even when \drea is unable to guarantee a single digit accuracy.
This improvement is more pronounced when half-precision arithmetic is used, as shown in~\cref{fig:dot_product_backward_error_half}.

\begin{figure}
    \centering
    \includegraphics[width=0.57\textwidth]{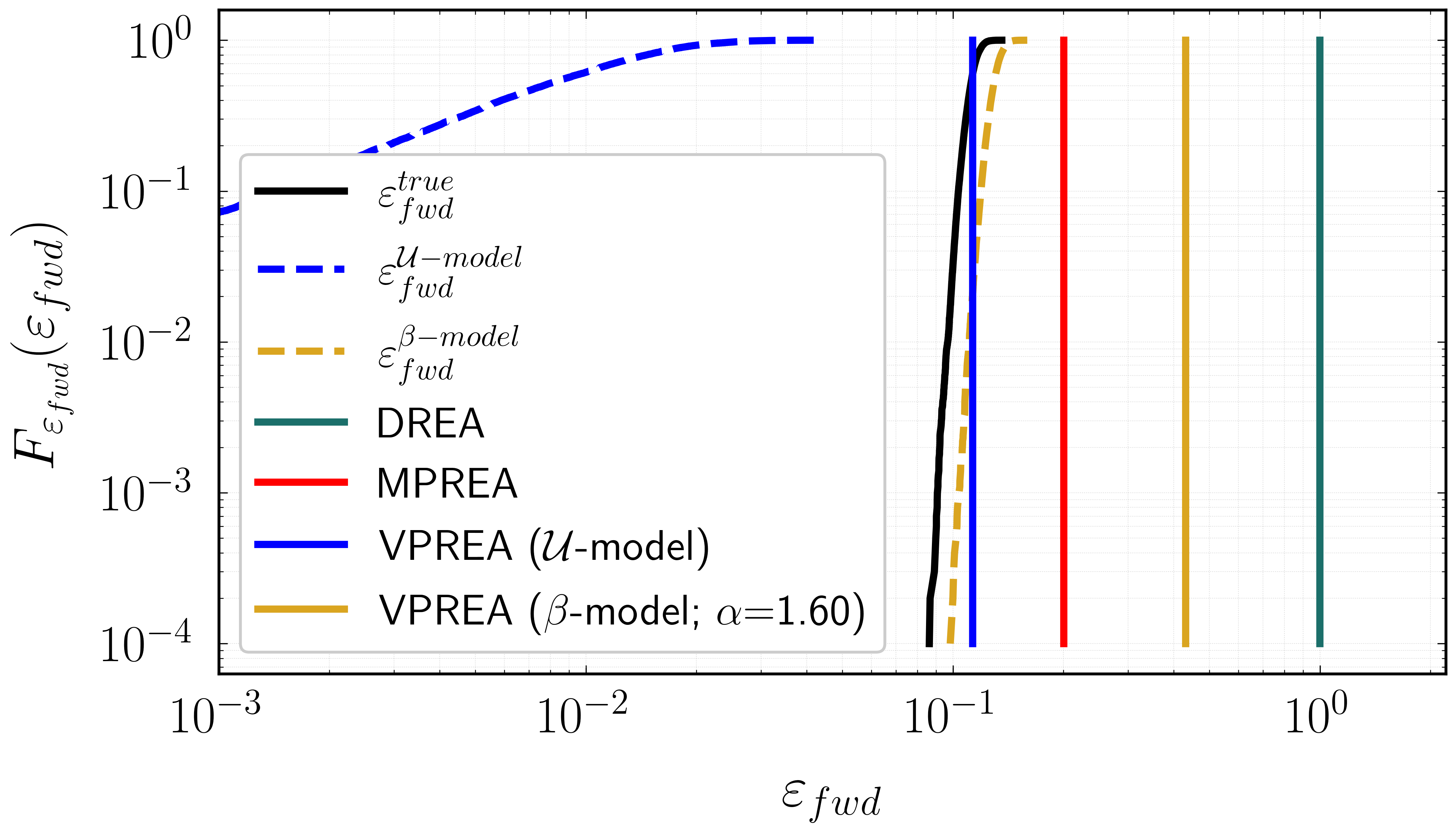}
    \caption{ Empirical distribution functions of the forward error in the dot product of random vectors of size $5\times 10^3$, with entries distributed as $\mathcal{U}(0,1)$, together with the modeled forward error (using~\cref{def:uniform_rounding_error_model,def:beta_rounding_error_model}) and the corresponding bounds.
    Computations are performed in half precision.
        All probabilistic bounds are evaluated at the confidence level
        $\mathcal{Q}(n;\zeta)=0.99$, with the $\beta$-model using shape parameter $\beta = 2.0$.
        Statistics are estimated from $10^4$ independent realizations.
        }
    \label{fig:dot_product_forward_error_half}
\end{figure}

\Cref{fig:dot_product_forward_error_half} illustrates the empirical distribution function (EDF) of the forward error, modeled forward error (using~\cref{def:uniform_rounding_error_model,def:beta_rounding_error_model}), and its bounds for the dot product of random vectors of size $5\times 10^3$ distributed as $\mathcal{U}(0,1)$ and computed using half-precision arithmetic.
For any random variable $X$, the EDF $F_X(x)$ is defined as $F_X(x) \define \frac{1}{N}\sum_{i=1}^{N} \mathbb{I}(X_i \leq x)$, where $N$ is the number of samples used to obtain the EDF and $\mathbb{I}$ is the indicator function.
As expected, the modeled forward error using the $\mathcal{U}$-model is unable to capture the forward error distribution accurately when the computations involve adding small positive numbers to large positive numbers, as discussed in~\cref{sec:methodology}.
However, the $\beta$-model more accurately captures the forward error distribution, and probabilistic bounds based on this model can estimate the forward error within an order of magnitude.
Note, despite \mprea providing tigher estimate for the forward error compared to \vprea($\beta$-model), it is an invalid bound as the $\expect{\delta}$ is non-zero, as discussed in~\cref{sec:methodology} and shown in~\cref{fig:rounding_error_distribution_small_increments}.

\subsection{Matrix-vector product}\label{sec:experiment_mv}
We now consider the matrix-vector product $\vect{y} = \mat{a}\vect{x}$, where $\mat{a}\in\real{n\times n}$ and $\vect{x}\in\real{n}$.
The backward error for this operation can be obtained using the Oettli-Prager theorem~\cite{oettli1964compatibility,higham2002accuracy} given by
\begin{align}
    \varepsilon_{bwd} = \min \left\{ \epsilon\geq 0: \hat{\vect{y}} = (\mat{a} + \Delta \mat{a})\vect{x}, \; \dfrac{\abs{\Delta \mat{a}}}{\abs{\mat{a}}} \leq \epsilon \right\} = \underset{i}{\max}\frac{\abs{\hat{\vect{y}} - \vect{y}}_i}{(\abs{\mat{a}}\abs{\vect{x}})_i},
\end{align}
where $\vect{\hat{y}}$ is the computed matrix-vector product using floating point arithmetic.
For the matrix $\mat{a}$, we consider real-valued square matrices with dimensions $n \leq 5\times 10^3$, selected to avoid overflow in half-precision arithmetic when $\norm[\infty]{\vect{x}} \leq 1$.
These matrices are drawn from the \software{SuiteSparse}~\cite{davis2011university} collection and comprise of $701$ problems arising from diverse real-world applications, including structural engineering, computational fluid dynamics, and chemical process simulation.
The matrices have different sparsity patterns, condition numbers, and numerical properties, providing a comprehensive testbed for evaluating the rounding error bounds.

\begin{figure}
    \centering
    \includegraphics[width=0.9\textwidth]{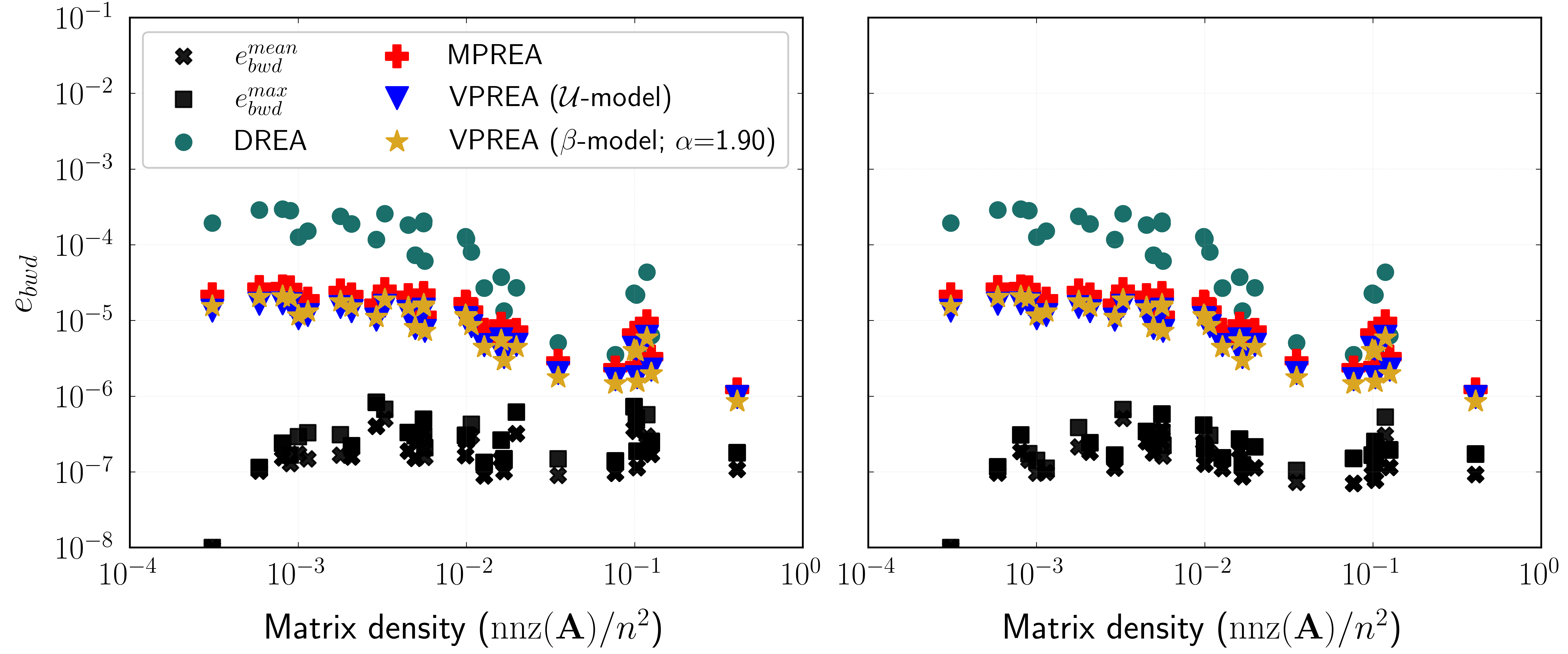}
    \caption{Backward error and its bounds for the matrix-vector product $\vect{y} = \mat{a}\vect{x}$, where $\vect{x}\ireal{n}$ is distributed as $\mathcal{U}(0,1)$ (left) and $\mathcal{U}(-1,1)$ (right), computed using single-precision arithmetic.
    Here, all matrices $\mat{a}\ireal{n\times n}$ are taken from the \software{SuiteSparse} collection~\cite{davis2011university}, and they contain $\mathrm{nnz}(\mat{A})$ non-zero entries.
    All probabilistic bounds are evaluated using a confidence level $\mathcal{Q}(n^2;\zeta) = 0.99$, and the $\beta$-model uses the shape parameter $\beta = 2.0$.
    To obtain the statistics, $10^2$ independent experiments were conducted for each matrix.
    }
    \label{fig:matrix_vector_product_backward_error_vs_nnz_single}
\end{figure}

\begin{figure}
    \centering
    \includegraphics[width=0.9\textwidth]{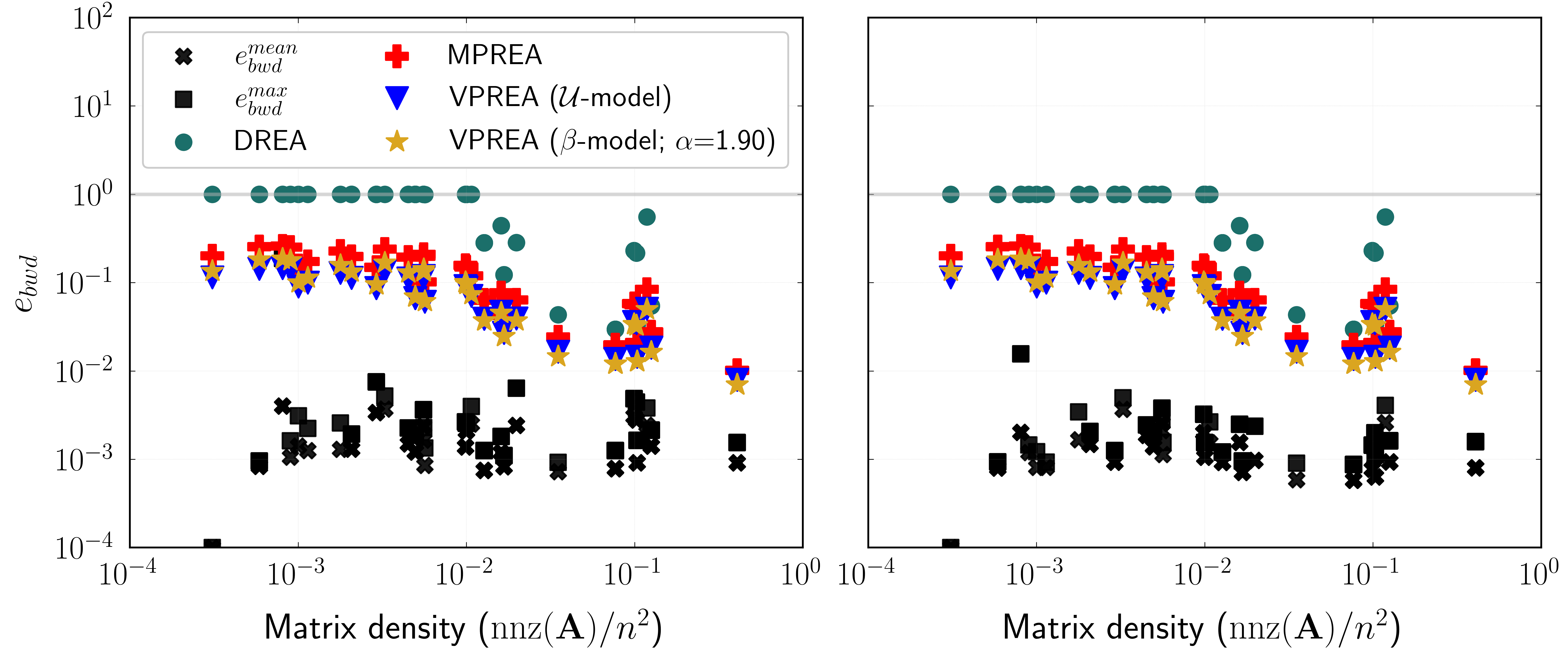}
    \caption{Backward error and its bounds for the matrix-vector product $\vect{y} = \mat{a}\vect{x}$, where $\vect{x}\ireal{n}$ is distributed as $\mathcal{U}(0,1)$ (left) and $\mathcal{U}(-1,1)$ (right), computed using half-precision arithmetic.
    Here, all matrices $\mat{a}\ireal{n\times n}$ are taken from the \software{SuiteSparse} collection~\cite{davis2011university}, and they contain $\mathrm{nnz}(\mat{A})$ non-zero entries.
    All probabilistic bounds are evaluated using a confidence level $\mathcal{Q}(n^2;\zeta) = 0.99$, and the $\beta$-model uses the shape parameter $\beta = 2.0$.
    To obtain the statistics, $10^2$ independent experiments were conducted for each matrix.
    All bounds are plotted until they exceed one (\textcolor{refgrey}{\fline}), beyond which they are not meaningful for backward error analysis.
    }
    \label{fig:matrix_vector_product_backward_error_vs_nnz_half}
\end{figure}

\Cref{fig:matrix_vector_product_backward_error_vs_nnz_single} illustrates the backward error and its bounds (using~\cref{theorem:vprea_matrix_vector}) for the matrix-vector product $\vect{y} = \mat{a}\vect{x}$, where $\vect{x}\ireal{n}$ is distributed as $\mathcal{U}(0,1)$ and $\mathcal{U}(-1,1)$ using single-precision arithmetic.
All probabilistic bounds yield nearly an order-of-magnitude improvement over deterministic bounds across a wide range of matrix density, parameterized by the fraction of non-zero entries present in the matrix.
As observed, the addition of the variance information in the probabilistic bounds only produced marginal improvements over the first-moment-based bound.
Further, for extremely sparse matrices, all bounds become quite pessimistic, as they ignore the sparsity pattern and the structure of the matrix.
We make a similar observation for the results obtained using half-precision arithmetic, as shown in~\cref{fig:matrix_vector_product_backward_error_vs_nnz_half}.
If known a priori, such sparsity information can be leveraged in the backward error analysis to obtain more accurate estimates.
For example, if the maximum number of non-zeros in any row is known to be $k_{\max}$, then we can trivially obtain the following result.
\begin{corollary}
    Let $\vect{y} = \mat{a}\vect{x}$, where $\mat{a}\ireal{m\times n}, \vect{x}\ireal{n}$ and $\vect{y}\ireal{m}$.
    Further, let the maximum number of non-zeros in any row of $\mat{a}$ be $k_{\max}$.
    Then, the computed matrix-vector product $\hat{\vect{y}}$ satisfies
    \begin{align*}
        \hat{\vect{y}} = (\mat{a} + \Delta \mat{a}) \vect{x},
    \end{align*}
    where $\abs{\Delta\mat{a}} \leq \hat{\gamma}_{k_{\max}}\abs{\mat{A}}$ holds true with a probability of at least $\mathcal{Q}(mk_{\max};\zeta)$.
    \begin{proof}
        The result follows directly from~\cref{theorem:vprea_matrix_vector} by noting that each row of $\mat{a}$ has at most $k_{\max}$ non-zero entries.
    \end{proof}
\end{corollary}


\subsection{Stochastic Boundary Value Problem}\label{sec:stochastic_bvp}
We now consider a boundary-value ordinary differential equation (ODE) with random coefficients and random forcing
\begin{equation}
    \frac{d}{d x} \Big( (1 + \theta_1 x) \frac{d u}{d x}\Big) = -50 \theta_2^2;\quad x\in [0, 1];~u(0) = u(1) = 0,
    \label{eq:ode}
\end{equation}
where $\theta_1 \sim \mathcal{U}(0.1, 1.1 )$ and $\theta_2 \sim \mathcal{U}( 1, 2 )$, similar to~\cite{giles2015multilevel}.
Here, $u: [0,1] \to \real{}$ is the solution function, and $\theta_1$ and $\theta_2$ are independent random variables representing model parameters.
The quantity of interest is $q = \expect{P}$, where $P$ is a random variable with a realization $p\define \int_x u\,dx$ which also has a closed-form expression given as
\begin{align}
    p(\theta_1, \theta_2) = \frac{25\theta_2^2(-2\theta_1 + (2 + \theta_1)\log(1 + \theta_1))}{\theta_1^2 \log(1 + \theta_1)}.
    \label{eq:realization_stochastic_bvp}
\end{align}

Numerically computing the quantity of interest $q$ is associated with multiple sources of uncertainty and numerical errors, namely, (a) parameter uncertainty, (b) Monte Carlo sampling uncertainty, (c) uncertainty due to floating-point arithmetic, and (d) numerical discretization error.
Here, we quantify the uncertainty arising from floating-point arithmetic in computing the quantity of interest, accounting for multiple sources of uncertainty and numerical errors.

\subsubsection{Finite-dimensional Approximation}
Computing the quantity of interest $q$ numerically first involves obtaining a finite-dimensional approximation of the ODE in~\cref{eq:ode} via numerical discretization.
For any given model parameters $\theta_1$ and $\theta_2$, the ODE system in~\cref{eq:ode} can be discretized via the finite-difference method using $M$ intervals of size $\Delta x \define \frac{1}{M}$.
Using second-order central difference approximation for the first- and second-order derivatives, the discretized form of~\cref{eq:ode} is given as
\begin{align}
    \Big(1 + \frac{\theta_1}{M}(i- \frac{1}{2})\Big) \vect{u}_{i-1}
    - 2\Big( 1 + \frac{\theta_1}{M}i\Big) \vect{u}_i
    + \Big(1 + \frac{\theta_1}{M}(i+\frac{1}{2})\Big) \vect{u}_{i+1} = -50\theta_2^2\Delta x^2,
    \label{eq:ode_discrete}
\end{align}
where $\vect{u}\in\real{M-1}$ is the discrete solution of $u(x)$ at $x=i\Delta x$ for $i=1,\dots, M-1$.
This results in a tri-diagonal system of equations $\mat{A}(\theta_1)\vect{u} = \vect{b}(\theta_2)$, where $\mat{A}\ireal{(M-1)\times (M-1)}, \vect{u}\ireal{(M-1)},$ and $\vect{b}\ireal{(M-1)}$.
Such a tri-diagonal system can be solved efficiently via the Thomas algorithm, as discussed in~\cref{sec:thomas}.
Once the discrete solution $\vect{u}$ is obtained, an approximation of the integral quantity $p$ can be obtained via Riemann integration as $p\approx\sum_{i=1}^{M-1} \vect{u}_i \Delta x$.
Lastly, the finite-dimensional approximation to the quantity of interest $q$ can be obtained via Monte Carlo integration using $N_s$ samples.

Obtaining an approximation for the realization $p$ using a finite-dimensional approximation $\vect{u}$ introduces numerical discretization error.
This error then propagates when computing an approximation to the quantity of interest $q$ using Monte Carlo integration.
Moreover, using finite samples of parameters to approximate the expectation $\expect{P}$ via Monte Carlo integration introduces sampling uncertainty.

\subsubsection{Discretization- and Sampling-Aware Computational Uncertainty due to Floating-Point Arithmetic}
Computing an approximation of the quantity of interest $q$ using finite-precision arithmetic results in several rounding errors.
Broadly, rounding errors accumulate when (a) obtaining the finite-dimensional numerical solution of~\cref{eq:ode} using a finite-difference scheme~\cref{eq:ode_discrete}, (b) performing Reimann integration using the obtained finite-dimensional solution of the tri-diagonal system of equations, and (c) computing the expectation $\expect{P}$ using finite realizations.
Given a tri-diagonal system of equations, here, we propagate the uncertainty due to floating-point arithmetic through each of the aforementioned computational steps.

\begin{figure}
    \centering
    \includegraphics[width=0.88\textwidth]{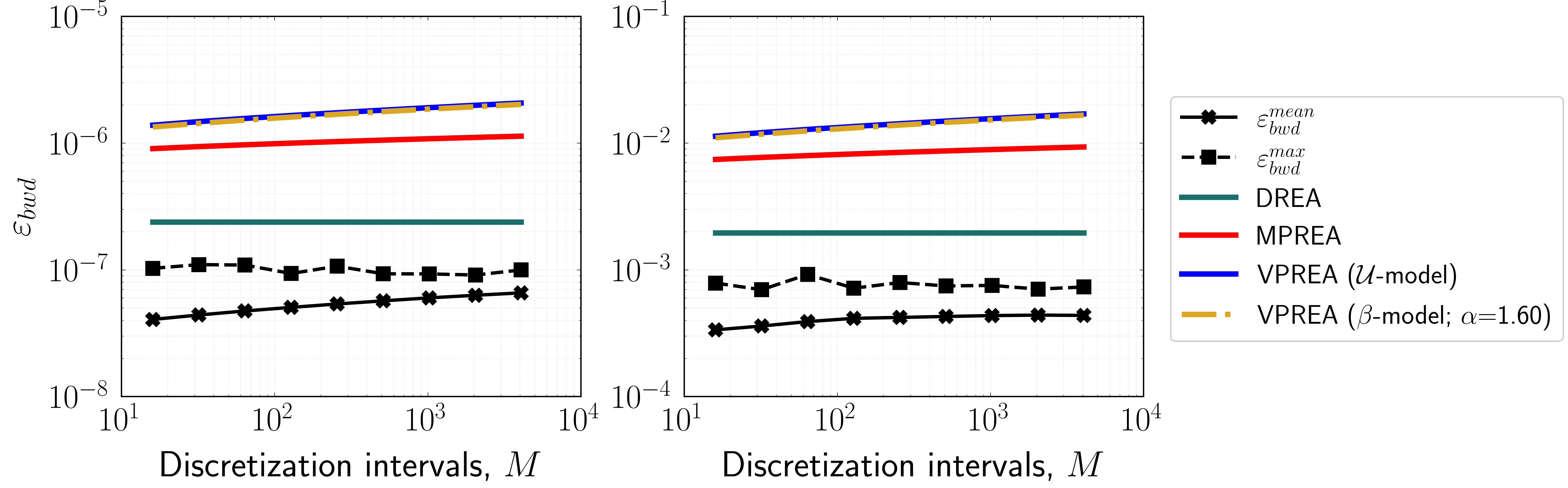}
    \caption{Backward error and its bounds for the solution of the tri-diagonal system of equations vs. number of discretization intervals $M$, computed in single-precision (left) and half-precision (right) floating-point arithmetic.
    All probabilistic bounds are evaluated using a confidence level $\mathcal{Q}(7(M-1);\zeta) = 0.99$, and the $\beta$-model uses the shape parameter $\beta = 2.0$.
    To obtain the statistics, $10^4$ independent experiments were conducted for each discretization interval.
    }
    \label{fig:bvp_backward_error}
\end{figure}
The backward error for solving the tri-diagonal system using the Thomas algorithm can be obtained via Oettli-Prager theorem~\cite[Theorem 7.3]{higham2002accuracy} as
\begin{align}
    \varepsilon_{bwd} = \min\left\{ \varepsilon\geq 0: (\mat{A} + \Delta \mat{A}) \hat{\vect{u}} = \vect{b},~\abs{\Delta \mat{A}}\leq \varepsilon \abs{\mat{a}}\right\} = \max_i\frac{\abs{ \mat{A}\hat{\vect{u}} -\vect{b}   }_i}{ (\abs{\mat{a}} \abs{\hat{\vect{u}}}    )_i},
   \label{eq:ode/lins_backward_error}
\end{align}
where $\hat{\vect{u}}$ is the computed solution of the linear system of equations in finite-precision arithmetic.
\Cref{fig:bvp_backward_error} presents the backward error and its bounds (using~\cref{theorem:vprea_thomas}) for solving the tri-diagonal system of equations using the Thomas algorithm in single- and half-precision floating-point arithmetic.
As observed, all bounds estimate the backward error within an order of magnitude.
However, for all discretization intervals $M$, \drea yields a tighter estimate than \mprea and \vprea.
This is expected as $\tilde{\gamma}_{n^\dagger} , \hat{\gamma}_{n^\dagger} \geq \gamma_{n^\dagger}$ for all $n^\dagger\lesssim 10$, as shown in~\cref{fig:gamma_vs_n}.

Using~\cref{theorem:vprea_thomas,theorem:vprea_dot_product}, we can now obtain that the computed approximation $\hat{p}$ of the integral quantity $p$ satisfies
\begin{align}
    \hat{p} = p + \Delta x\sum_{i=1}^{M-1} \Delta \vect{u}_i,
\end{align}
where $\Delta \vect{u}_i$ is the $i$-th component of the perturbation $\Delta \vect{u}$ due to floating-point arithmetic propagated from the solution of the Thomas algorithm to the Riemann integral.
As a result, $\Delta \vect{u}$ satisfies $\abs{\Delta \vect{u}}\leq \hat{\gamma}_{M-1}\left(\abs{\hat{\vect{u}}} + \left(2\hat{\gamma}_1 + \hat{\gamma}_2 + \hat{\gamma}_1\hat{\gamma}_2\right)\abs{\mat{a}^{-1}}\abs{\mat{a}}\abs{\hat{\vect{u}}}\right)$ that holds true with a probability of at least $\mathcal{Q}(7(M-1)^2 - 5(M-1);\zeta)$ via principle of inclusion and exclusion.
Now, we can obtain that the computed approximation $\hat{q}$ of the quantity of interest $q$ using $N_s$ samples satisfies
\begin{align}\label{eq:stochastic_bvp_qoi_error}
    \hat{q} = q + \frac{1}{N_s}\sum_{k=1}^{N_s} \Delta p_k,
\end{align}
where $\Delta p_k$ is the error in computing the $k$-th realization of the random variable $P$ that satisfies
\begin{align*}
    \abs{\Delta p_k} \leq \hat{\gamma}_{N_s}\left(\abs{\hat{p}_k} + \Delta x\sum_{i=1}^{M-1}\abs{\Delta\vect{u}_i}\right),
\end{align*}
with a probability of at least $\mathcal{Q}(N_s(7(M-1)^2 - 5(M-1) + 1);\zeta)$.
Using~\cref{eq:stochastic_bvp_qoi_error}, we obtain discretization- and sampling-aware probabilistic bounds for quantifying the uncertainty induced by floating-point arithmetic in the computation of the quantity of interest $q$.
These bounds depend only on the computed solution $\hat{\vect{u}}$ of the tri-diagonal linear system and the computed Riemann integral $\hat{p}$, and do not require access to the exact solution of the ODE in~\cref{eq:ode}.
\begin{figure}
    \centering
    \includegraphics[width=0.85\textwidth]{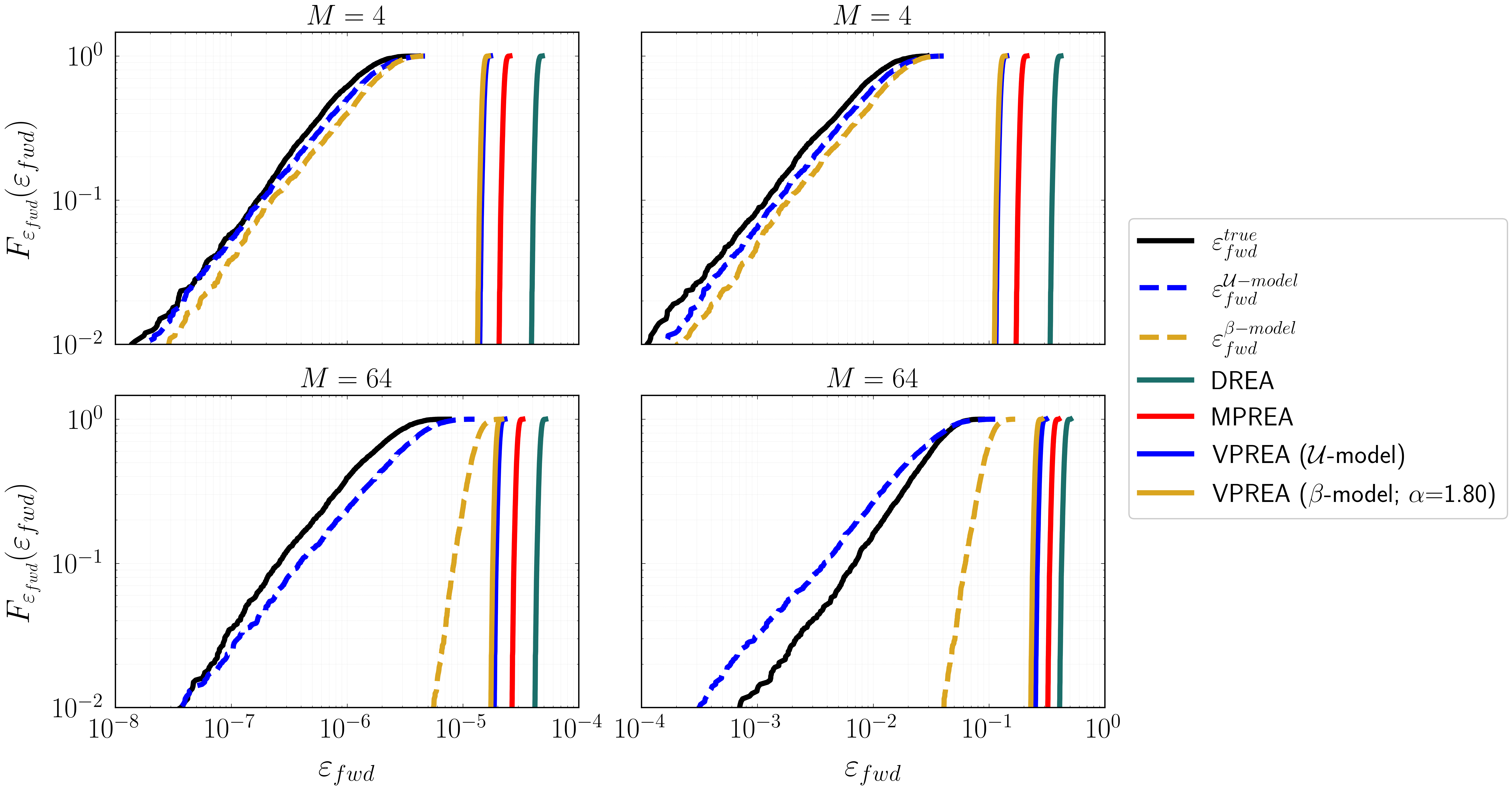}
    \caption{Empirical distribution functions of the forward error in the quantity of interest $q$, along with the modeled forward error (using~\cref{def:uniform_rounding_error_model,def:beta_rounding_error_model}) and the corresponding bounds.
    Results are shown for $N_s = 100$ Monte Carlo samples, computed in single precision (left column) and half precision (right column).
        All probabilistic bounds are evaluated at the confidence level
        $\mathcal{Q}\!\left(N_s(7(M-1)^2 - 5(M-1) + 1);\zeta\right)=0.99$,
        with the $\beta$-model using shape parameter $\beta = 2.0$.
        Statistics are estimated from $2\times10^3$ independent realizations for each discretization interval $M$.
        }
    \label{fig:bvp_forward_error_num_samples_100}
\end{figure}

\begin{figure}
    \centering
    \includegraphics[width=0.85\textwidth]{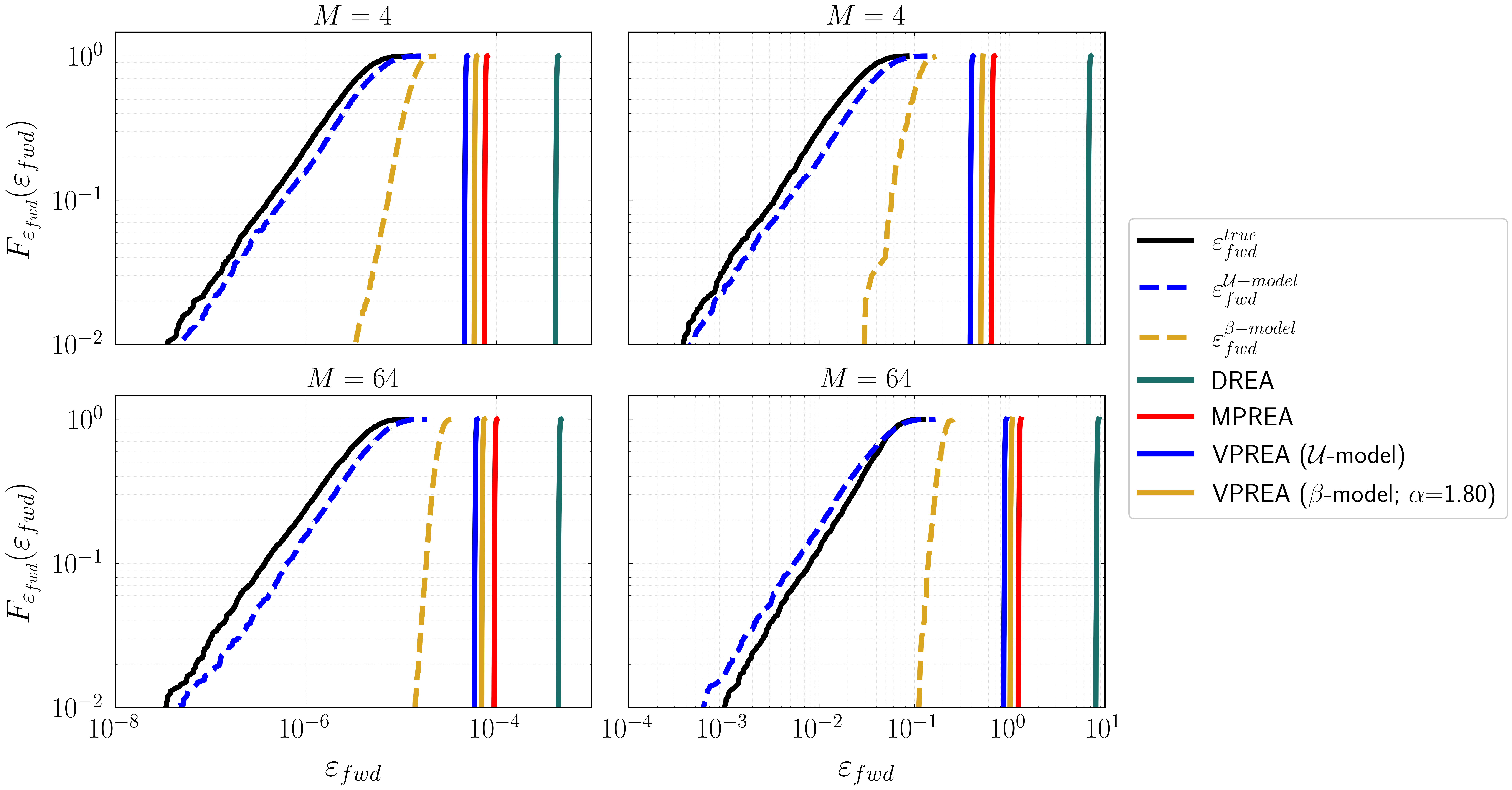}
    \caption{Empirical distribution functions of the forward error in the quantity of interest $q$, along with the modeled forward error (using~\cref{def:uniform_rounding_error_model,def:beta_rounding_error_model}) and the corresponding bounds.
    Results are shown for $N_s = 1000$ Monte Carlo samples, computed in single precision (left column) and half precision (right column).
        All probabilistic bounds are evaluated at the confidence level
        $\mathcal{Q}\!\left(N_s(7(M-1)^2 - 5(M-1) + 1);\zeta\right)=0.99$,
        with the $\beta$-model using shape parameter $\beta = 2.0$.
        Statistics are estimated from $2\times10^3$ independent realizations for each discretization interval $M$.
        }
    \label{fig:bvp_forward_error_num_samples_1000}
\end{figure}

\Cref{fig:bvp_forward_error_num_samples_100} presents the EDF of the forward error, modeled forward error (using~\cref{def:uniform_rounding_error_model,def:beta_rounding_error_model}), and its bounds for computing $q$ using $N_s = 100$ Monte Carlo samples.
Both the $\mathcal{U}$-model and the $\beta$-model accurately approximate the forward error distribution, and the resulting probabilistic bounds yield tighter estimates than deterministic bounds.
While the deterministic bounds yield relatively tight estimates for the solution of the tri-diagonal linear system (see~\cref{fig:bvp_backward_error}), they do not adequately reflect the increment in the floating-point uncertainty as the number of arithmetic operations increases.
In particular, the uncertainty due to floating-point errors grows during the Riemann integration (with increasing discretization intervals $M$) and during the Monte Carlo integration (with increasing number of samples $N_s$).
While both deterministic and probabilistic approaches capture this progressive accumulation, the probabilistic bounds characterize it much more sharply, whereas the deterministic bounds become increasingly conservative.
\Cref{fig:bvp_forward_error_num_samples_1000} shows the corresponding results for $N_s = 1000$.
As the number of Monte Carlo samples increases, the deterministic bounds become more pessimistic.
In contrast, probabilistic bounds provide an order-of-magnitude improvement over deterministic bounds and continue to yield tight estimates of the forward error distribution.

\section{Concluding remarks}\label{sec:conclusion}
In this work, we strengthen the probabilistic backward error framework of Higham and Mary~\cite{higham2019new} by introducing \emph{Variance-informed Probabilistic Rounding Error Analysis} (\vprea). The proposed framework defines a confidence-calibrated, operation-count-dependent constant $\hat{\gamma}_n$ that incorporates distributional structure beyond the classical zero-mean assumption.
The analysis relies only on the first two moments of the rounding error random variable, characterized in log-space, and avoids higher-order moment assumptions. Specifically, we leverage Bernstein's concentration inequality to analyze the logarithm of multiplicative rounding error terms, transforming their product into a sum of random variables and enabling probabilistic bounds on error accumulation.
Under the assumption that rounding errors are independent and identically distributed, \vprea\ yields exact backward error bounds valid for arbitrary operation counts and precision levels, providing a systematic framework for quantifying floating-point uncertainty in modern low-precision computing environments.

A central contribution is the explicit characterization of bias in rounding error propagation.
We introduce two parametric models for the rounding error distribution: a $\mathcal{U}$-model, which assumes a uniform distribution and recovers the classical zero-mean setting, and a $\beta$-model, in which rounding errors follow a $\operatorname{Beta}$ distribution that enables controlled introduction of bias.
We derive explicit conditions under which positive or negative bias can be introduced within \vprea\ through the $\beta$-model.

Importantly, we show that the growth of the probabilistic constant depends on how the rounding error distribution is parameterized.
Our analysis shows that distributional parametrization can alter the growth of probabilistic bounds; in particular, under the $\beta$-model, transient growth can transition from $\mathcal{O}(\sqrt{n})$ to $\mathcal{O}(n)$.
Using the dot product as a canonical example, we demonstrate that bias-aware modeling yields tight bounds in both half and single precision arithmetic, accurately capturing observed accumulation across precisions.

Numerical experiments on sparse matrix-vector products show that accounting for sparsity is essential for obtaining meaningful bounds, leading to a corollary that incorporates sparsity into the backward error analysis. For a stochastic boundary value problem, we develop discretization- and sampling-aware probabilistic bounds that quantify floating-point uncertainty alongside discretization and statistical errors; the resulting bounds improve classical deterministic guarantees by nearly an order of magnitude. 

Overall, these results that probabilistic rounding error bound growth is not intrinsic but depends on how rounding errors are modeled. By making this dependence explicit, \vprea\ provides a principled framework for tighter, confidence-calibrated error bounds in large-scale, low-precision scientific computing.

\section*{Reproducibility of computational results}
The code for all experiments is publicly available at \url{https://github.com/sahilbhola14/FinUQ}.

\section*{Acknowledgments}
This work is dedicated to the memory of the late Prof. Nick Higham whose work continues to inspire the field of  numerical analysis.
We thank Profs. Jean-Baptiste Jeannin and Alex Gorodetsky for advice and for the valuable feedback on the early drafts of this manuscript.
This research was supported by NSF Grant FMITF-2219997.

\bibliographystyle{siamplain}
\bibliography{references}
\end{document}